\documentclass[prb,twocolumn,longbibliography,amsmath,floatfix,superscriptaddress,amssymb,nobibnotes]{revtex4-2}

\usepackage{tikz}
\usepackage{makeidx,amssymb,amsmath,amsthm,graphicx}
\usepackage[toc,page]{appendix}
\usepackage{dcolumn}
\usepackage{bm}
\usepackage{color}
\usepackage{placeins}
\usepackage{epstopdf}
\usepackage{multirow}
\usepackage{rcs}
\usepackage[normalem]{ulem}
\usepackage{chngcntr}
\usepackage[thinspace,thinqspace]{SIunits}
\usepackage{natbib}
\usepackage[linkcolor=blue,urlcolor=blue,citecolor=blue,colorlinks=true]{hyperref}

\usepackage[T1]{fontenc}            
\usepackage[latin1]{inputenc}

\usepackage{babel}
\makeatother
\begin{document}


\title{Single crystal growth and properties of Au- and Ge-substituted EuPd$_2$Si$_2$} 
\author{Michelle~Ocker}
\affiliation{%
 Kristall- und Materiallabor, Physikalisches Institut, 
 Goethe-Universit\"at Frankfurt, Max-von-Laue Stasse 1, 
 60438 Frankfurt am Main, Germany
}%
\author{Robert~M\"oller}
\affiliation{%
 Kristall- und Materiallabor, Physikalisches Institut, 
 Goethe-Universit\"at Frankfurt, Max-von-Laue Stasse 1, 
 60438 Frankfurt am Main, Germany
}%
\author{Marius~Peters}
\affiliation{%
 Kristall- und Materiallabor, Physikalisches Institut, 
 Goethe-Universit\"at Frankfurt, Max-von-Laue Stasse 1, 
 60438 Frankfurt am Main, Germany
}%
\author{Franziska~Walther}
\affiliation{%
 Kristall- und Materiallabor, Physikalisches Institut, 
 Goethe-Universit\"at Frankfurt, Max-von-Laue Stasse 1, 
 60438 Frankfurt am Main, Germany
}%
\author{Vivien~Kirschall}
\affiliation{%
 Geowissenschaften, 
 Goethe-Universit\"at Frankfurt, Altenh\"ofer Allee, 
 60438 Frankfurt am Main, Germany
}%
\author{Dominik C.~Hezel}
\affiliation{%
 Geowissenschaften, 
 Goethe-Universit\"at Frankfurt, Altenh\"ofer Allee, 
 60438 Frankfurt am Main, Germany
}%
\author{Michael Merz}
\email[Corresponding author:]{michael.merz@kit.edu}
\affiliation{Institute for Quantum Materials and Technologies, Karlsruhe Institute of Technology, Kaiserstr. 12, 76131 Karlsruhe, Germany}
\affiliation{Karlsruhe Nano Micro Facility (KNMFi), Karlsruhe Institute of Technology, Kaiserstr. 12, 76131 Karlsruhe, Germany}
\author{Christo Guguschev}
\affiliation{Leibniz-Institut f\"ur Kristallz\"uchtung, Max-Born-Str. 2, 12489 Berlin, Germany
}

\author{Cornelius~Krellner}
\affiliation{%
 Kristall- und Materiallabor, Physikalisches Institut, 
 Goethe-Universit\"at Frankfurt, Max-von-Laue Stasse 1, 
 60438 Frankfurt am Main, Germany
}%
\author{Kristin~Kliemt}
\email[Corresponding author:]{kliemt@physik.uni-frankfurt.de}
\affiliation{%
 Kristall- und Materiallabor, Physikalisches Institut, 
 Goethe-Universit\"at Frankfurt, Max-von-Laue Stasse 1, 
 60438 Frankfurt am Main, Germany
}%

\date{\today}

\begin{abstract}
We report on the single crystal growth of Eu(Pd$_{1-x}$Au$_x$)$_2$Si$_2$, $0< x\leq 0.2$, from a levitating Eu-rich melt using the Czochralski method. Our structural analysis of the samples confirms the ThCr$_2$Si$_2$-type structure as well as an increase of the room temperature $a$ and $c$ lattice parameters with increasing $x$. Chemical analysis reveals that, depending on the Au concentration, only about 25-35\% of the amount of Au available in the initial melt is incorporated into the crystal structure, resulting in a decreasing substitution level for increasing $x$. Through Au substitution, chemical pressure is applied and large changes in valence crossover temperatures are already observed for low substitution levels $x$.  In contrast to previous studies, we do not find any signs of a first-order transition in samples with $x_{\rm nom}=0.1$ or AFM order for higher $x$. Furthermore, we observe the formation of quarternary side phases for a higher amount of Au in the melt. \\
In addition, cubic-mm-sized single crystals of EuPd$_2$(Si$_{1-x}$Ge$_x$)$_2$ with $x_{\rm nom}=0.2$ were grown. The analysis of the X-ray fluorescence revealed that the crystals exhibit a slight variation in the Ge content. Such tiny compositional changes can cause changes in the sample properties concerning variations of the crossover temperature or changes of the type of the transition from crossover to magnetic order.
Furthermore, we report on a new orthorhombic phase EuPd$_{1.42}$Si$_{1.27}$Ge$_{0.31}$ that orders antiferromagnetically below $17\,\rm K$.

\end{abstract}


\maketitle

\section{\label{sec:level1}Introduction}
In recent years, Eu-based materials have become the  focus of research in solid state physics because they are suited for the study of a variety of different phenomena such as topological states \cite{Ma2019}, colossal magnetoresistance \cite{Wang2021, Krebber2023}, intermediate valence \cite{Bauminger1978, Sampathkumaran1981}, skyrmion phases \cite{Onuki2020}, or critical elasticity \cite{Wolf2022, Wolf2023}.  
In compounds, Eu can occur in different valence states. Most of the known Eu-based materials contain magnetic Eu$^{2+}$ ions (4$f^7$ configuration with spin momentum S = 7/2, orbital momentum L = 0, total angular momentum J = 7/2) and show magnetic order \cite{Feng2010, Hedo2014}. Some compounds contain non-magnetic Eu$^{3+}$ ions (4$f^6$ configuration with S = 3, L = 3, J = 0) \cite{seiro2013anomalous} and do not order magnetically to the lowest temperatures. There are some examples \cite{Bauminger1978, Sampathkumaran1981} in which Eu shows intermediate valence and undergoes a valence transition when changing temperature or pressure. $p-T$ phase diagrams of different materials have been published, providing a systematic overview of their overall pressure dependence \cite{Onuki2017, Seiro2011, Walther2025}. 
An outstanding and versatile material is  EuPd$_2$Si$_2$. 
Upon application of external \cite{Vijakumar1981} as well as chemical pressure \cite{Mitsuda1999, Cho2002, Peters2023, Segre1982, Croft1982} changes in the lattice and physical properties are huge \cite{Batlogg1982} making it a suitable material for studying critical elasticity and effects of a strong electron-lattice coupling \cite{Wolf2022, Wolf2023}. 
\\
In EuPd$_2$Si$_2$, the valence-crossover temperature $T^{\prime}_V\approx 160\,\rm K$ can be reduced by isoelectronic substitution of Ge at the Si site and the system undergoes a transition from intermediate valent to a stable Eu$^{2+}$ configuration for $x_{\rm nom}\geq0.2$ exhibiting antiferromagnetic order with $T_N=47\,\rm K$ \cite{Peters2023}. 
Here, we present the analysis of a large Ge-substituted Czochralski-grown sample with $x_{\rm nom}=0.2$. A concentration close to where the intermediate-valent to magnetic order transition occurs. There is evidence that the $x=0.105$ Ge substituted sample undergoes a first-order phase transition from antiferromagnetic order to a valence crossover under the application of a small He-gas pressure of only $p=0.1\,\rm GPa$, and will allow further study of critical elasticity in the vicinity of the critical endpoint \cite{Wolf2023}. 

The Czochralski-grown EuPd$_2$(Si$_{1-x}$Ge$_x$)$_2$ sample with $x_{\rm nom}=0.2$ is characterized with respect to growth behavior, appearance of side phases, and disorder. In addition, we describe the structure, composition, and magnetic properties of a new phase in the Eu-Pd-Si-Ge system, which occurs as a side phase in our growth experiments. \\
In the case of Au substitution, the changes in physical properties are already huge for small Au substitution levels \cite{Croft1982,Segre1982}, meaning that the substitution-induced disorder is expected to be smaller compared to Ge-substitution. 
In the past, the series Eu(Pd$_{1-x}$Au$_x$)$_2$Si$_2$ was studied using polycrystalline material due to the lack of single crystals \cite{Croft1982,Segre1982,Gupta1982, Abd-Elmeguid1985_ZfPB}. In susceptibility data, Eu$^{2+}$-impurity contributions were usually observed at low temperatures \cite{Gupta1982}.
A complex phase diagram was developed from the results of $^{151}$Eu M\"ossbauer and susceptibility measurements. It shows magnetic order for $0.175\leq x^{\rm poly}\leq0.25$, first-order valence transitions between $0.05<x^{\rm poly}<0.175$ terminating at a critical endpoint (CEP) at $x^{\rm poly}\approx 0.04$ and a valence crossover regime between $x^{\rm poly}\approx0.04$ and the unsubstituted EuPd$_2$Si$_2$ \cite{Croft1982, Segre1982}. 
The M\"ossbauer isomer shift versus temperature determined for a sample with $x^{\rm poly}=0.15$ as well as its dependence on $x$ at $12\,\rm K$ show first-order jumps. In these samples showing the first-order transition, the transition is smeared out because of substitution inhomogeneities of the crystals, resulting in the transition of different crystal regions at slightly different temperatures. The authors report that the samples for their study crystallize in the ThCr$_2$Si$_2$ structure but without showing any chemical analysis \cite{Croft1982, Segre1982}. 
The authors also mentioned that the antiferromagnetic sample with $x^{\rm poly}=0.25$ has a slightly enhanced valence of $\approx 2.18$ determined by L$_{\rm III}$ absorption spectroscopy \cite{Segre1982} which posed the question whether an intermediate valent compound might order magnetically or not. 
In later studies, it was found that the valence deduced from the L$_{\rm III}$ X-ray absorption measurements cannot be directly compared with the M\"ossbauer isomer shift and that the Eu valence becomes unstable for $0.18\leq x^{\rm poly}\leq 0.4$ where the system shows AFM order, while Eu is stable 2+ for higher $x$ \cite{Abd-Elmeguid1985_ZfPB, Abd-Elmeguid1985a}. In their samples, the authors found a non-linear decrease in the $a$-lattice parameter as a function of Au substitution, but confirmed that all samples exhibit the ThCr$_2$Si$_2$ structure \cite{Abd-Elmeguid1985a, Abd-Elmeguid1985_ZfPB}.
The proposed phase diagram in Refs.~\cite{Croft1982, Segre1982} was later called into question by another M\"ossbauer spectroscopy study, where the results were interpreted as not showing a first-order transition for samples with $x^{\rm poly}<0.175$ \cite{Sauer1987}. Subsequent X-ray diffraction results \cite{Jhans1988} again support the existence of a first-order transition in Eu(Pd$_{1-x}$Au$_x$)$_2$Si$_2$.\\
To gain further insight in this controversial issue and to explore the potential of the system with regard to critical elasticity in more detail, we have grown Au-substituted EuPd$_2$Si$_2$ single-crystalline samples by the Czochralski method from a levitating melt in this study. 
After careful structural and chemical characterization of the samples, we explored the effect of the increasing Au substitution on the valence transition in the magnetic susceptibility as well as in the heat capacity. 
The questions to answer are whether samples with low Au concentrations show a first-order valence transition and if so, at which $x$ the critical endpoint is reached. Samples close to this particular concentration would be suitable candidates for studying the expected strong effects of electron-lattice coupling.

\begin{figure}
\centering
\includegraphics[width=0.35\textwidth]{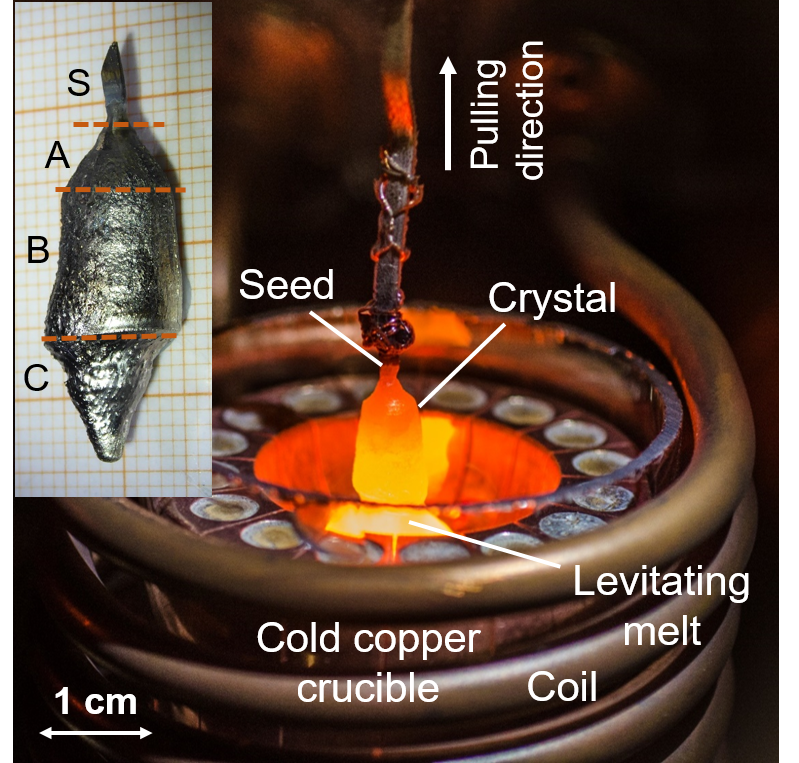}
	\caption[]{Czochralski growth from a levitating melt. {\it Inset:} Result of the Czochralski growth process of a sample EuPd$_2$(Si$_{1-x}$Ge$_x$)$_2$ with a nominal Ge concentration of $x = 0.20$ on a mm-grid. S marks the seed crystal, A the area where the target phase crystallizes, and B the part where inclusions of a secondary phase appear. In part C, the side  phase EuPd$_{1.42}$Si$_{1.27}$Ge$_{0.31}$ occurs exclusively.}
\label{sample-MP809_Czochralski}
\end{figure}
\begin{figure}
\centering
\includegraphics[width=1.0\linewidth]{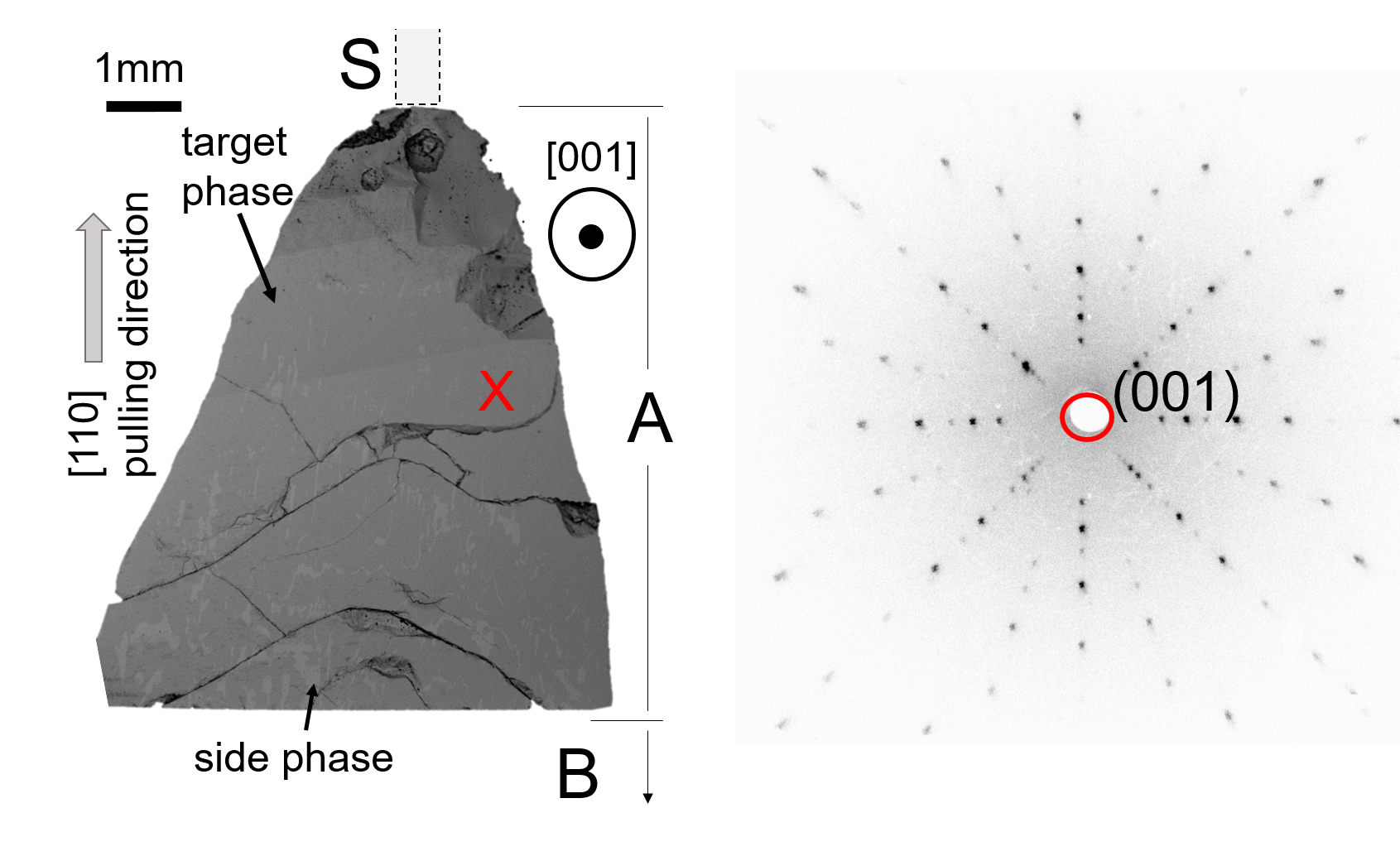}
	\caption[]{Eu(Pd$_{1-x}$Au$_x$)$_2$Si$_2$, $x_{\rm nom}=0.1$ (a) SEM image of an oriented cut and polished single crystal. The red cross marks the location where the Laue diffraction pattern presented in (b) was recorded. The sharp spots in the diffraction pattern indicate the high crystallinity of the grown sample.}
\label{SEM_Au-subst_EuPd2Si2}
\end{figure}
\begin{figure}[b]
\centering
\includegraphics[width=1.0\linewidth]{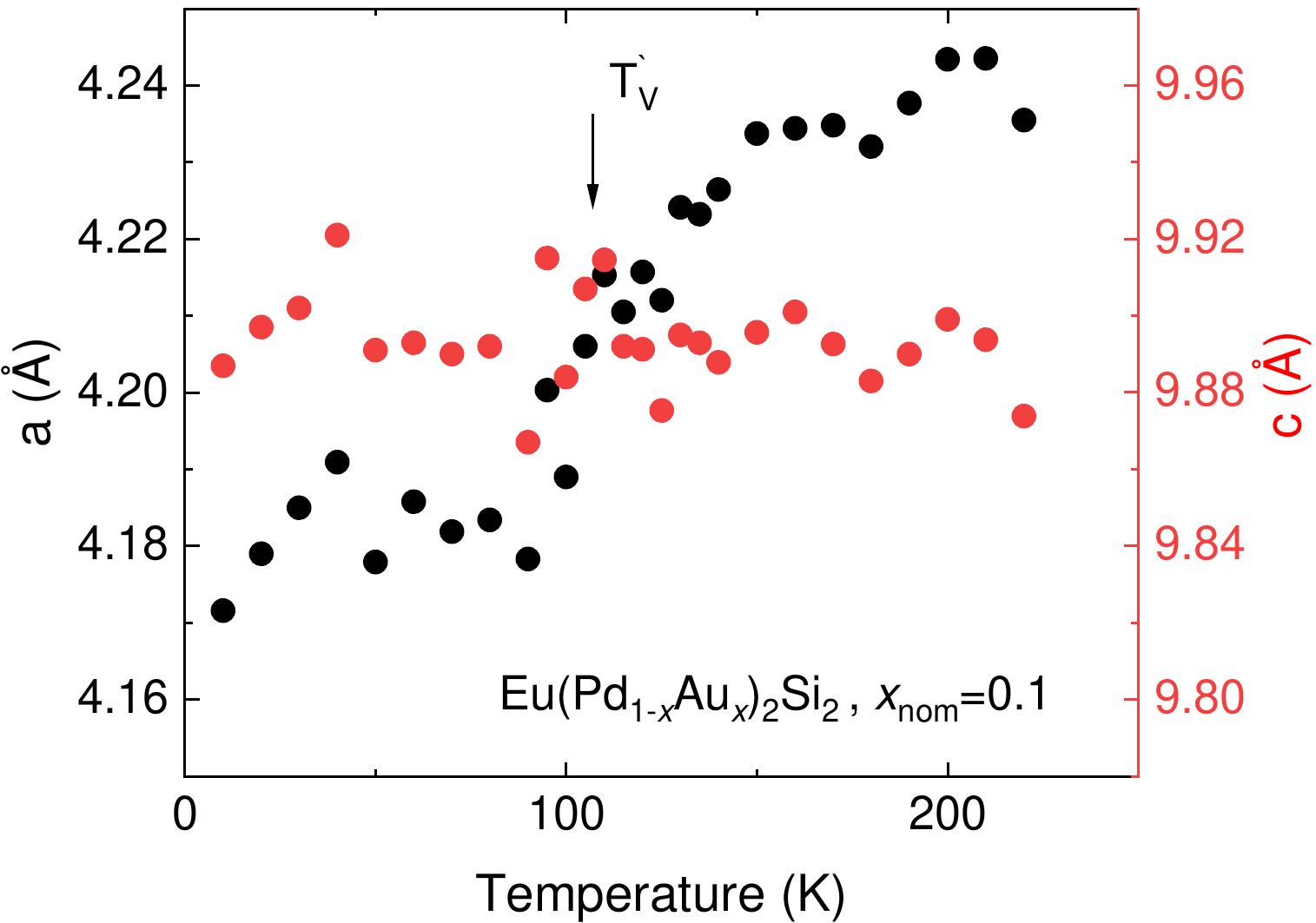}
	\caption[]{Temperature dependence of the lattice parameters of Eu(Pd$_{1-x}$Au$_x$)$_2$Si$_2$, $x_{\rm nom}=0.1$. Corresponding PXRD data are shown in Fig.~\ref{TT-PXRD_Au-subst_EuPd2Si2}.}
\label{TTPXRD_Lattice_Au-subst_EuPd2Si2}
\end{figure}

\section{Experimental Details}
High purity elements Eu (99.99\%, chunks, Alpha Aesar), Pd (99.999\%, rod, Heraeus), Si (99.9999\%, pieces, Cerac), Ge (99.9999\%, pieces, Otavi Minen), Au (99.99\%) were used for all crystal growth experiments. The growth experiments, Fig.~\ref{sample-MP809_Czochralski}, of EuPd$_2$(Si$_{1-x}$Ge$_x$)$_2$ with a nominal Ge content of $x_{\rm nom}=0.20$, and Eu(Pd$_{1-x}$Au$_x$)$_2$Si$_2$ with a nominal Au content of $x_{\rm nom}=0.10, 0.15$, and $0.20$ in the initial melt were performed from a levitating Eu-rich melt using initial weights of 15\,g.  
For the crystal growth, we used the Czochralski method under Ar overpressure ($p=20\,\rm bar$) in analogy to the procedure described in \cite{Kliemt2022a, Peters2023}. The initial stoichiometry and the solidus temperatures of $T_{\rm sol}\approx1150 ^{\circ}$C determined pyrometrically at the beginning of each growth experiment are shown in Tab.~\ref{tab:3} in \cite{SI_Ausubst2025}.
A typical growth result is presented in Fig.~\ref{SEM_Au-subst_EuPd2Si2}(a). The samples were mapped to find single-crystalline pieces, and the orientation of the single crystals was determined using a Laue camera with white x-rays of a tungsten anode. Laue patterns were simulated using the software OrientExpress \cite{orientexpress}. 
Energy-dispersive x-ray spectroscopy (EDX) in combination with scanning electron microscopy (SEM) using a  Zeiss DSM 940 A, with an additional energy-dispersive detector (EDAX Ametek GmbH), was utilized to determine the chemical composition and to search for inclusions of secondary phases. 
Element compositions were determined by wavelenght-dispersive x-ray spectroscopy (WDX) using a JEOL 8530F Plus Hyperprobe.
The accelerating voltage was set to 15\,kV with a beam current of 20\,nA. The following elements were measured: Si, Pd, Au, Ge, Eu. The spot analyses were performed with a focused beam of 2 $\mu$m diameter. Peak measurement times were between 20 and 30\,s, and backgrounds were measured with half peak measurement times. Pure metal standards, and in case of Eu a well characterised phosphate were used for calibration and the build-in PRZ ARMSTRONG correction was applied. The standards were calibrated to <1 rel\%, except for Au, which has $\approx 5$ rel\%. Detection limits were - with wt-ppm in brackets: Si (150), Pd (700), Au (4000), Ge (300), Eu (1300).\\
Micro X-ray fluorescence ($\mu$-XRF) elemental mapping and energy dispersive Laue mapping (EDLM) were carried out using a Bruker M4 TORNADO spectrometer to: (1) investigate the chemical homogeneity of cleaved and polished sections prepared from the grown crystals and (2) assess the microstructure. 
EDLM has proven capabilities to detect low- and high-angle grain boundaries, twins, striations, and dislocations in crystalline materials \cite{guguschev2015microstructural,subramanian2023investigation, guguschev2022revisiting, guguschev2024application}. The measurement system was equipped with a rhodium X-ray source operated at 50\,kV and and 600\,$\mu$A. Polycapillary X-ray optics were used to focus the non-polarized white radiation at the surface of the sample, resulting in a spatial resolution of about 20\,$\mu$m. The white radiation from the excitation source interacts with the crystals, and because of the instrument geometry, it is possible to detect Bragg reflections. The principles of the XRD surface mapping technique, the measurement procedure, and the measurement setup are described elsewhere \cite{guguschev2015microstructural}. 
The analysis of the sample areas for the polished and cleaved samples was done at a pressure of 20 and 1000\,mbar, respectively. The step size between the spot measurements was 10 and 30\,$\mu$m. The surface areas were scanned "on the fly" by moving the sample stage continuously for 3 to 8 cycles. The measurement time per point was set between 20 and 35\,ms and two silicon drift detectors (SDD) were used.\\ 
Powder X-ray diffraction (PXRD) in a Bruker D8 diffractometer and low temperature powder x-ray diffraction (LT-PXRD) in a Siemens D500 diffractometer using Cu-K$_{\alpha}$ radiation were performed to verify the structure of the crushed single crystals and determine the lattice parameters. 
For the refinement of the lattice parameters, PowderCell 2.3 software \cite{PowderCell} and GSAS II \cite{Toby2013} were used. 
The crystal structure was investigated through single-crystal x-ray diffraction (SC-XRD). Data were collected using a high-flux, high-resolution Rigaku Synergy-DW diffractometer equipped with a rotating anode (Mo/Ag) and utilizing Mo $K_\mathrm{\alpha}$ radiation ($\lambda = 0.7107$ \AA). The instrument features precisely aligned Montel mirror optics, a motorized divergence slit set to 5 mrad, and a background-less Hypix-Arc150$^\circ$ detector, ensuring minimal reflection profile distortion and consistent detection of all reflections under equivalent conditions. The investigated specimen was measured at 300 K in shutterless 32-bit mode to a resolution better than 0.4\,{\AA} and displayed no mosaic spread or reflections from secondary phases, indicating very high sample quality. These features allowed for precise data collection and reduction using the latest version of the CrysAlisPro software package \cite{CrysAlis}. The crystal structure was solved and refined using JANA2006 \cite{Vaclav_229_2014}, incorporating all symmetry-independent reflections with intensities greater than 2$\sigma$. The unit cell parameters and space group were determined, and atomic positions were initially localized using random phases methods. The structure was then completed and refined through difference Fourier synthesis. The refinement converged very well, yielding excellent residuals (see Tables \ref{XRDtable1} and \ref{XRDtable2} below, including $wR_\mathrm{2}$, $R_\mathrm{1}$, and goodness-of-fit (GOF) values). 
Heat capacity and magnetization measurements were carried out using the commercial measurement options of a Quantum Design PPMS.

\section{Results}
The preparation of Eu-containing materials is difficult as a result of the high reactivity of Eu with oxygen and its high vapor pressure. However, in the recent past, progress has been made in the growth of single crystals of different Eu compounds \cite{Onuki2017, Onuki2020, Jo2020, Kliemt2022a, Peters2023, Wang2021, Krebber2023, Usachov2024}.
In addition, developing the crystal growth procedures for intermetallic Eu-based systems is often quite elaborate, since in most cases no ternary phase diagrams or isothermal sections are available in the data bases. 
From the few known examples \cite{ASM_Babizhetskii1997, ASM_Kalychak1998} it is obvious that the formation of a variety of different neighboring phases for a given compound can be expected.

\subsection{Absence of first-order transitions in Eu(Pd$_{1-x}$Au$_x$)$_2$Si$_2$}

\subsubsection{Crystal Growth}\noindent
The samples obtained in the crystal growth experiments of the series Eu(Pd$_{1-x}$Au$_x$)$_2$Si$_2$ have an appearance similar to that shown in Fig.~\ref{sample-MP809_Czochralski}. A SEM image of a polished cut through the upper part A of a sample is depicted in Fig.~\ref{SEM_Au-subst_EuPd2Si2}(a), where the gray area S marks the position of the seed crystal, which came from unsubstituted EuPd$_2$Si$_2$. In area A the target phase (gray) formed without or with minor amount of inclusions of a secondary phase (light gray). Part B marks the area in which a secondary phase occurred in a higher fraction. Despite the fact that single crystalline EuPd$_2$Si$_2$ seed crystals were used, samples grown to several mm long often consisted of more than one single crystal grain. The extraction of single-crystalline samples from the grown ingot for the physical characterization was done using the Laue method and a spark erosion device or by cleaving the samples. Fig.~\ref{SEM_Au-subst_EuPd2Si2}(a) shows a cut through a large single crystalline sample with $x_{\rm nom}=0.1$ perpendicular to the crystallographic $[001]$ direction. The red cross marks the location where the Laue pattern in Fig.~\ref{SEM_Au-subst_EuPd2Si2}(b) was recorded. The materials considered here form incongruently and a change in melt composition occurs when they are grown.  By pulling the crystal out, the residual melt enriches with Eu and Au in the course of the growth, which leads to the formation of side phases.
When the sample was cooled to room temperature after growth, several cracks developed (dark lines in Fig.~\ref{SEM_Au-subst_EuPd2Si2}(a)) as a result of different thermal expansion of the included side phase and the target phase.

\subsubsection{Powder X-ray diffraction }
\begin{table*}[htbp]
\begin{center}
\begin{tabular}{|c|c|c|c|c|c|c|c|}
\hline
$x_{\rm nom}$&$x_{\rm WDX}$   &$a$&$c$&$V$ &$T^{\prime}_v$ HC&$T^{\prime}_v$ VSM&Ref.\\
             &  & [\AA] &[\AA]  &[\AA$^3$]   & [K]   & [K]              & \\
\hline
0                           & 0&4.2396(5) &9.8626(4)&177.3&154&154& \cite{Kliemt2022a}\\
0.1           &0.028(4) & 4.2456(6) & 9.8663(4) &177.8&102-119&100&this work\\
0.15          &0.046(5)& 4.2452(7)  &9.8693(3)&177.9&77-78&77-78&this work\\
0.2           &0.065(5)&4.2506(4)   &9.8738(5)&178.4&90&89&this work\\
\hline
\end{tabular}
\end{center}
\caption{Eu(Pd$_{1-x}$Au$_x$)$_2$Si$_2$ with different $x_{\rm nom}$: Au substitution level $x_{\rm WDX}$ as determined by WDX, room-temperature lattice parameters and valence crossover temperatures.}
\label{tab:1}
\end{table*}

Through the refinement of the PXRD data shown in Fig.~\ref{PXRD_Au-subst_EuPd2Si2} in the SI \cite{SI_Ausubst2025}, we found that for increasing Au substitution, the ThCr$_2$Si$_2$ structure is maintained and the room-temperature lattice parameters $a$ and $c$ as well as the unit cell volume increases as shown in Tab.~\ref{tab:1} and Fig.~\ref{Lattice_Au-subst_EuPd2Si2} in the SI \cite{SI_Ausubst2025} which is consistent with data presented in Ref.~\cite{Sauer1987}.
In the growth with $x_{\rm nom}=0.2$, there was a high amount of a quarternary side phase. This side phase was isolated and its diffraction pattern is shown in Fig.~\ref{sidephase_0.2_RM102_PXRD} \cite{SI_Ausubst2025}. 

The volume of the unit cell changes strongly when crossing the valence transition since the atomic radii of the larger Eu$^{2+}$ (1.17\,\AA) and the smaller Eu$^{3+}$ (0.95\,\AA) differ significally \cite{Mueller2008}. 
The temperature dependence of the lattice parameters of a sample with $x_{\rm nom}=0.1$ which was investigated using LT-PXRD is shown in Fig.~\ref{TTPXRD_Lattice_Au-subst_EuPd2Si2}. The data were obtained through Rietveld refinement of the LT-PXRD data in  Fig.~\ref{TT-PXRD_Au-subst_EuPd2Si2} \cite{SI_Ausubst2025}. Upon cooling, the $c$ parameter stays nearly constant while $a$ undergoes a contraction of $\approx$ 2\% when crossing $T_v^{\prime}$. The inflection point of the data indicates $T_v^{\prime}\approx 110\,\rm K$ which is consistent with the $T_v^{\prime}$ determined from HC and VSM data for this substitution level presented below. The relative change of the $a$ parameter is comparable to that observed for pure and Ge-substituted samples \cite{Peters2023}.

\subsubsection{WDX, EDX and Au incorporation}

The Au concentration $x_{\rm WDX}$ of the crystals was determined using WDX and EDX. In Tab.~\ref{tab:1}, the results of the WDX analysis concerning the Au incorporation are shown for the different nominal Au concentrations $x_{\rm nom}$.
We found that the Au incorporation rate, e.g. the fraction of Au in the initial melt that is built into the crystal, decreases for higher $x_{\rm nom}$ and is between 35\% and 25\%. This is much lower than the incorporation rate of 70\% to 50\% for Ge in the series EuPd$_2$(Si$_{1-x}$Ge$_x$)$_2$ \cite{Peters2023}. The WDX results are comparable to the EDX data, Tab.~\ref{tab:WDX} in \cite{SI_Ausubst2025}, but the error is larger in the latter, as no standard was used here. 
The element ratios of the target phase for different Au concentrations $x_{\rm nom}$ are shown in Tab.~\ref{tab:WDX} in \cite{SI_Ausubst2025}, and reveal constant Eu and Si contents, as well as decreasing Pd together with increasing Au content. This is consistent with an Au substitution at the Pd site. From the EDX and WDX analysis, it is obvious that the length of the sample that can be grown in the target phase depends on the Au substitution level. For the $x_{\rm nom}=0.1$ concentration, the sample consisted of target phase to an amount of more than 90\% in the first 10 mm below the seed with minor inclusions of secondary phases. The length of the target phase reduced from 3\,mm (with purity $\approx$ 95\%) for the $x_{\rm nom}=0.15$ Au content to only 1\,mm for the $x_{\rm nom}=0.2$ sample (with purity $\approx$ 80\%). 
The crystals for the magnetization and heat capacity measurements were chosen such that the side-phase content was as small as possible. However, for the highest substitution level $x_{\rm nom} = 0.2$ we could not extract a phase pure Eu(Pd$_{1-x}$Au$_x$)$_2$Si$_2$ crystal and the side phase contribution becomes rather dominant in the magnetization data.\\
As recently reported, the WDX analysis revealed a slight change in the Si-Pd ratio along the grown crystal for unsubstituted EuPd$_2$Si$_2$  \cite{Kliemt2022a}. Here, we observe a small change in this ratio as well as depicted in Fig.~\ref{WDX_0p15_Au-subst_EuPd2Si2} in \cite{SI_Ausubst2025} within one sample with $x_{\rm nom}=0.15$. Note that the Eu and the Au contents are constant throughout the length of the grown sample where the target phase was found. In the grown ingot, a side phase appeared below $\approx 3\,\rm mm$.\\
In Tab.~\ref{tab:WDX_sidephase} in \cite{SI_Ausubst2025}, the result of the side-phase analysis is shown. In the side phases, the Eu and Au content is higher, while the Si and Pd content is lower than in the 122 target phase. It is observed that the side phase stoichiometry is similar in the case of $x_{\rm nom}=0.1$ and $x_{\rm nom}=0.15$ ($\approx$ Eu$_{26}$Pd$_{33.5}$Si$_{34}$Au$_{6.5}$) while a slightly different side phase stoichiometry is found in the case $x_{\rm nom}=0.2$ ($\approx$ Eu$_{25}$Pd$_{30}$Si$_{35}$Au$_{10}$). 

\subsubsection{Heat capacity}

\begin{figure}
\centering
\includegraphics[width=1.0\linewidth]{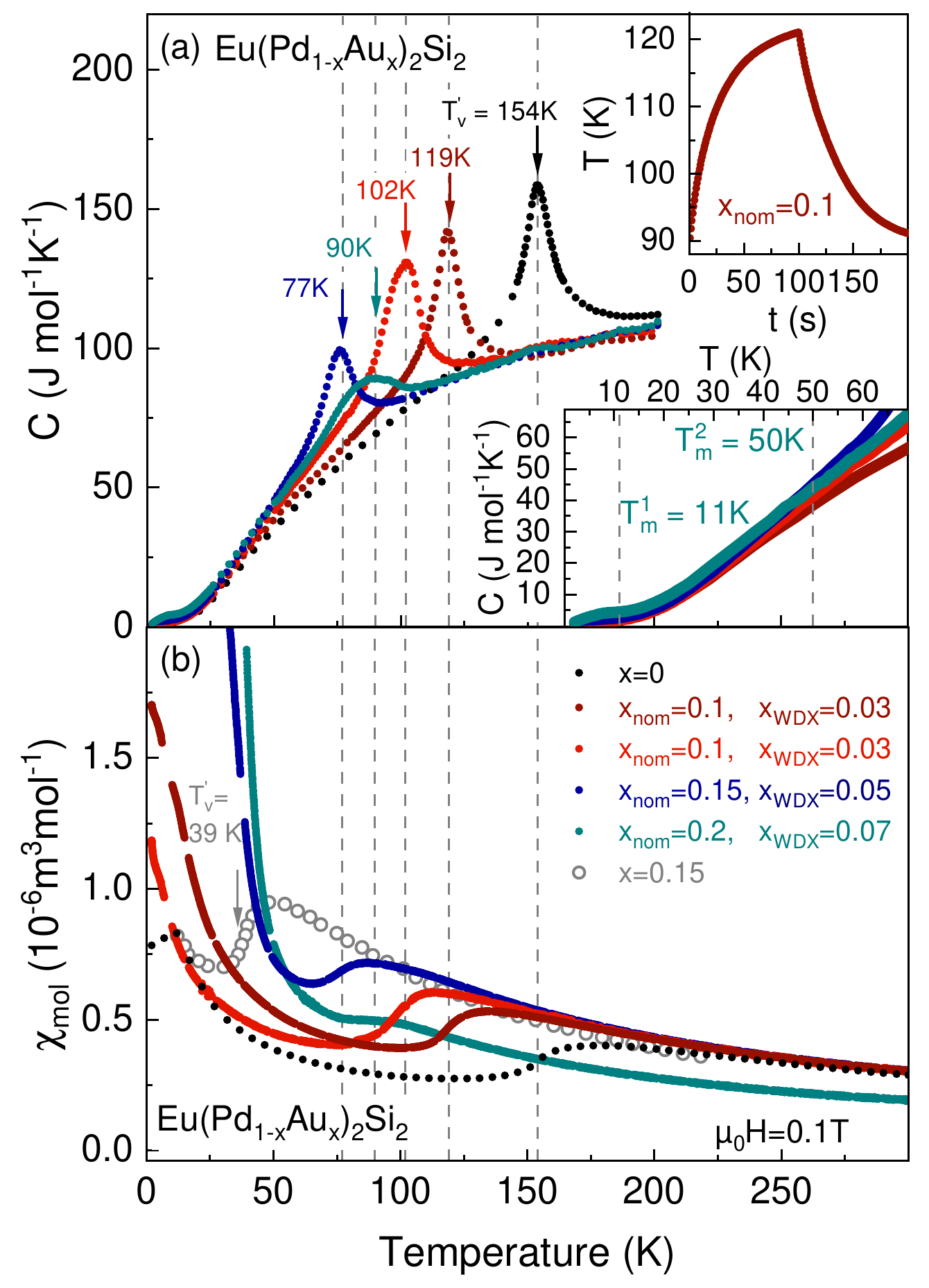}
	\caption[]{Eu(Pd$_{1-x}$Au$_x$)$_2$Si$_2$ (a) Temperature dependence of the heat capacity, $x=0$ data were taken from Ref.~\cite{Peters2023}. The upper inset shows the time dependence of a long heat pulse for $x=0.1$. The lower inset shows an enlarged view of the data below 70\,K. $T_m^1$, and $T_m^2$ mark the temperatures where magnetic transitions of a side phase appear. (b) Temperature dependence of the magnetic susceptibility in $\mu_0H=0.1\,\rm T$. The data for $x=0$ were taken from \cite{Kliemt2022a} and $x=0.15$ (gray) from \cite{Segre1982}. Arrows mark the crossover temperatures $T^\prime_v$ for the samples with the different Au concentrations.} 
\label{HC_Au-subst_EuPd2Si2_v1}
\label{ChivT_Au-subst_EuPd2Si2_red}
\end{figure}\noindent
The heat capacity (HC) measured between 1.8\,K and 200\,K is presented in Fig.~\ref{HC_Au-subst_EuPd2Si2_v1}. 
At the valence crossover temperature $T^{\prime}_v$, a peak in heat capacity occurs. 
All crossover temperatures that were determined for Au-substituted EuPd$_2$Si$_2$ are summarized in Tab.~\ref{tab:1}. Our data clearly show that with an increase of the Au content up to $x_{\rm nom}=0.15$ (blue curve in Fig.~\ref{HC_Au-subst_EuPd2Si2_v1}), $T^{\prime}_v$ decreases. According to data from the literature \cite{Croft1982, Segre1982}, a first-order transition was detected for samples with $0.05<x<0.175$. Therefore, we investigate the sample with $x_{\rm nom}=0.1$ (red and dark red circles in Fig.~\ref{HC_Au-subst_EuPd2Si2_v1}) in more detail by measuring the heat capacity of several samples extracted from different locations in the large Czochralski-grown sample. By comparing the valence-crossover temperatures, we found that samples originating from a site closer to the seed of the crystal show a lower $T^{\prime}_v=102\,\rm K$ (red curve in Fig.~\ref{HC_Au-subst_EuPd2Si2_v1}, sample extracted $\approx 1\,\rm mm$ below the seed), while those stemming from a site closer to the flux show a higher $T^{\prime}_v=119\,\rm K$ (dark red curve in Fig.~\ref{HC_Au-subst_EuPd2Si2_v1}, sample extracted $\approx 9\,\rm mm$ below the seed). This shift in $T^{\prime}_v$ is most likely caused by a slight change in the Pd-Si ratio and is consistent with previous observations for unsubstituted EuPd$_2$Si$_2$ \cite{Kliemt2022a}. 
The HC shows no evidence for the occurrence of a first-order transition for the samples in the series. At a valence transition that is of first order, a sharp peak in the HC is expected. We find that for $x_{\rm nom}\leq0.15$, the characteristics of the peak connected to the valence crossover do not change significantly. Our study of the crossover using large heat pulses, inset of Fig.~\ref{HC_Au-subst_EuPd2Si2_v1}, does not show the occurrence of latent heat for any of the samples which would be an indicator for a first-order transition. Instead of a sharpening of the peak, we observe a broadening for some samples, indicating the presence of a slight variation of the stoichiometry in one sample. 
This slight phase separation could not be resolved in our chemical analysis and is within the error of the WDX measurements, as given in Tab.~\ref{tab:WDX} in \cite{SI_Ausubst2025}. 
Compared to the sample with $x_{\rm nom}=0.1$, the height of the peak in the HC of that with $x_{\rm nom}=0.15$ is reduced and it is shifted to $T^{\prime}_v=77\,\rm K$. 
For $x_{\rm nom}=0.2$ (teal curve in Fig.~\ref{HC_Au-subst_EuPd2Si2_v1}), the peak broadens and occurs at a slightly higher temperature at $T_v=90\,\rm K$ than for $x_{\rm nom}=0.15$. In addition to this peak, we observe a change in slope at $T^1_m=50\,\rm K$ and a small hump in the HC of this sample at $T^2_m=11\,\rm K$, inset of Fig.~\ref{HC_Au-subst_EuPd2Si2_v1}, which are most likely connected to magnetic transitions of side phases. Fig.~\ref{HC_Au-subst_EuPd2Si2_RM102b} in \cite{SI_Ausubst2025} shows the analysis of a sample in more detail that was obtained in the growth of $x_{\rm nom}=0.2$, and consists mainly of the side phase. From this analysis, we can clearly assign the transition at $T^1_m$ and $T^2_m$ to a side phase.

\subsubsection{Magnetic susceptibility}\noindent
In Fig.~\ref{ChivT_Au-subst_EuPd2Si2_red}, we present the data of the magnetic susceptibility, $\chi(T)$, at $\mu_0H=0.1\,\rm T$ as a function of temperature measured upon cooling between 300~K and 2~K for the different Au concentrations.
The temperature of the valence crossover, $T^{\prime}_v$, was determined from the maximum of d$(\chi(T)\cdot T)/$d$T$ which corresponds to the inflection point of the magnetic susceptibility (for details, see \cite{Wolf2023}).
Although the valence crossover can be clearly seen in the $x_{\rm nom}=0.1$ data (red and dark red), it is smeared out for $x_{\rm nom}=0.15$ (blue) and can hardly be detected for $x_{\rm nom}=0.2$ (teal). 
Upon increasing the Au substitution level to $x_{\rm nom}=0.15$, the crossover temperature decreases to $T^{\prime}_v=78\,\rm K$ which is much higher than the $T^{\prime}_v=39\,\rm K$ found for polycrystalline samples with $x_{\rm nom}=0.15$ (gray curve in Fig.~\ref{ChivT_Au-subst_EuPd2Si2_red}) \cite{Segre1982}. 
For $x_{\rm nom}=0.2$, $T^{\prime}_v$ is difficult to detect accurately due to the high side-phase contribution. We found that $T^{\prime}_v$ increases again to $\approx 89\,\rm K$ which is consistent with the peak appearing in the HC data for this concentration. Note that when measuring upon cooling and heating, a hysteresis can be detected whose size depends on the sweep rate and disappears for sufficiently low sweep rate during measurements. In previous work \cite{Croft1982, Segre1982, wada1997temperature}, a steep increase in $\chi(T)$ was taken as an indication of a first-order transition. 
The large $x_{\rm nom}=0.1$ crystal had a low side-phase content, and we studied two small samples from this growth extracted from different parts of the large sample. The slight change in the Si-Pd ratio along the sample causes a shift in $T^{\prime}_v$ that is visible in the magnetic susceptibility similar to the heat capacity, and we display two data sets in Fig.~\ref{ChivT_Au-subst_EuPd2Si2_red}. 
We observe the same trend as previously reported for unsubstituted EuPd$_2$Si$_2$ \cite{Kliemt2022a} where a lower crossover temperature was found for samples extracted close to the seed. 

In the past, the low-temperature increase in susceptibility was assigned to impurities in polycrystalline samples  of Eu(Pd$_{1-x}$Au$_x$)$_2$Si$_2$ \cite{Gupta1982}.
In the $\chi(T)$ data of our samples, contributions of magnetic impurities and side phases show up below $T=50\,\rm K$. For $x_{\rm nom}=0.15$ we observe at least two magnetic transitions at $T_m^1=11\,\rm K$, and $T_m^2=50\,\rm K$. And for $x_{\rm nom}=0.2$, the magnetic transition at $T_m^2$ becomes dominant. 
Although the data for $x_{\rm nom}\leq0.15$ in Fig.~\ref{ChivT_Au-subst_EuPd2Si2_red} show comparable absolute values at $T>200~\rm K$, the data for $x_{\rm nom}=0.2$ do not follow the same trend, hinting at a significant fraction of the quaternary side phase in this sample. This also supports the fact that the magnetic transition at $T_m^2=50\,\rm K$ observed here is not intrinsically connected to Au substituted EuPd$_2$Si$_2$, but is a property of the side phase.

\subsection{EuPd$_2$(Si$_{1-x}$Ge$_x$)$_2$ with $x_{\rm nom}=0.2$ at the edge of valence crossover}
In previous work \cite{Peters2023}, we observed that some samples grown from a melt with $x_{\rm nom}=0.2$ show double transitions or varying magnetic properties. We have now grown and investigated  additional EuPd$_2$(Si$_{1-x}$Ge$_x$)$_2$ samples to gain insight on how homogeneous the samples of this most interesting substitution value are. 
Fig.~\ref{sample-MP809_Czochralski} shows the result of a Czochralski growth experiment of EuPd$_2$(Si$_{1-x}$Ge$_x$)$_2$ with a Ge content in the melt of $x_{\rm nom}=0.2$.

\subsubsection{Energy dispersive and non-dispersive Laue method}
\begin{figure}
\centering
\includegraphics[width=0.5\textwidth]{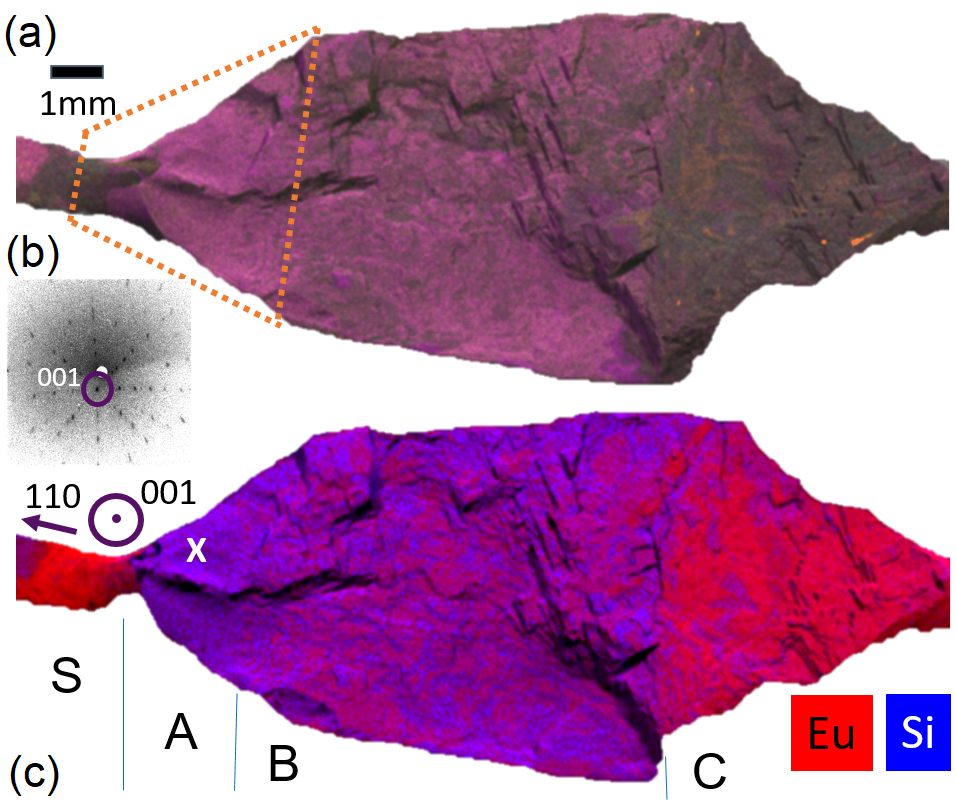}
	\caption[]{EuPd$_2$(Si$_{1-x}$Ge$_x$)$_2$, $x_{\rm nom}=0.2$ with cleaved surface; (a) The EDLM image  shows one large area of the same violet color indicating the presence of one single crystal grain in part A. 
    (b) Laue image of the crystallographic $[001]$ direction recorded in part A (white cross). The pulling direction of the crystal was along $[110]$.
    (c) The $\mu$-XRF spectroscopy image shows the elemental distributions of Eu and Si. In part A, the homogeneous blue-violet color indicates a constant Eu-Si ratio while in part B Eu-rich phases (red-violet) occur. }
\label{EDLM_all}
\label{RFA_all}
\end{figure}

\noindent
The energy dispersive Laue method (EDLM) can be used to detect the existence of SC grains with different orientations in large Czochralski-grown samples.
Each Bragg peak intensity is color-coded and five different Bragg peaks were selected to superimpose their 2D color-coded intensity distribution maps. The appearance of different colors in such maps typically indicates different grains or compounds. 
Fig.~\ref{EDLM_all}(a) shows the EDLM image of the cleaved sample from Fig.~\ref{sample-MP809_Czochralski}. From the picture it becomes clear that on this cleaved surface the orientation along one crystallographic direction of the target phase (violet) is maintained from the seed over the areas A and B.
The brownish and orange regions in B and C indicate the appearance of different orientations and/or side phases in these regions.

The cleaved surface of the target phase in area A exhibits orientation (001) as determined by the Laue method, Fig.~\ref{EDLM_all}(b).

\begin{figure}
\centering
\includegraphics[width=0.5\textwidth]{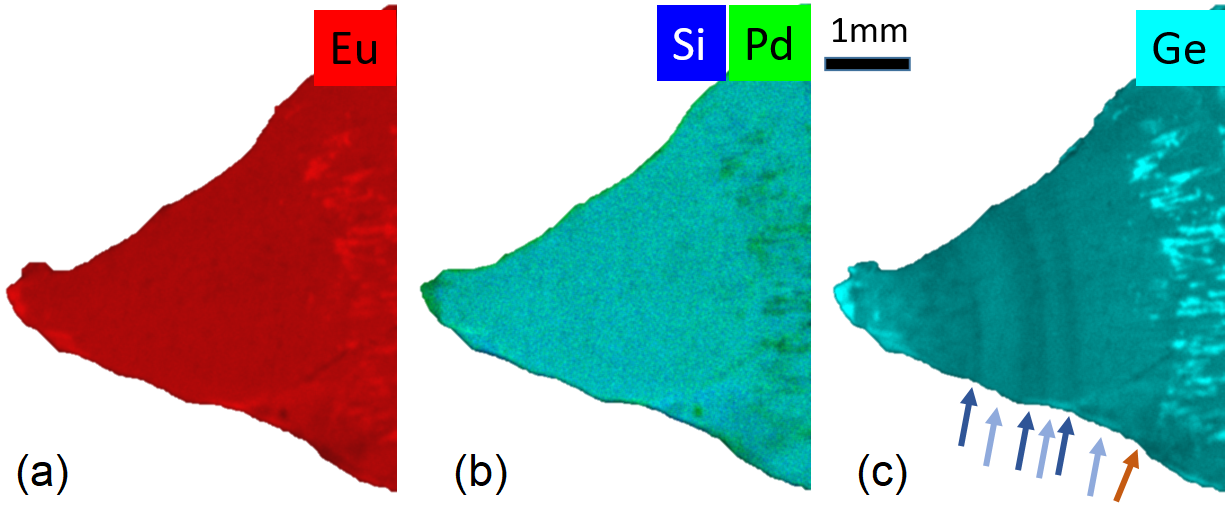}
	\caption[]{EuPd$_2$(Si$_{1-x}$Ge$_x$)$_2$, $x_{\rm nom}=0.2$: Polished surface of the sample that is shown in Fig.~\ref{RFA_all}  (orange box, area A). $\mu$-XRF spectroscopy: Element distribution images for (a) Eu, (b) Pd, Si and (c) Ge. }
\label{elements}
\end{figure}

\subsubsection{Micro X-ray fluorescence spectroscopy }
Micro X-ray fluorescence spectroscopy ($\mu$-XRF spectroscopy) detects elemental distributions of for instance Eu and Si, Fig.~\ref{RFA_all}(c) at the cleaved surface. While in part A of the sample, the occurrence of a homogeneous blue-violet color indicates a constant Eu-Si ratio, in part B, Eu-rich phases (red-violet) start to emerge. Part C is dominated by a Eu-rich side phase. After polishing, the elemental distribution in part A of the same sample was examined in more detail to get insights into the growth mechanism. 
In Fig.~\ref{elements}, this upper part A is shown. The seed was attached to the tip visible on the left hand side of the sample. The target phase was formed starting from the seed which is visible as a region of homogeneous color indicating homogeneous elemental distribution of Eu in Fig.~\ref{elements}(a) and of Pd-Si in Fig.~\ref{elements}(b). In Fig.~\ref{elements}(c), the Ge distribution is shown which exhibits clear striations perpendicular to the growth direction indicating slight changes of the Ge content. The regions with lower (higher) Ge content are marked by dark (light) blue arrows. 
After $\approx 3.5\,\rm mm$ (orange arrow), a strong change of composition is visible and Eu- and 
a Ge-rich phase grows together with the target phase. The shape of the convex interface is visible by the striations and by the interface where the secondary phases started to appear.

\subsubsection{Powder X-ray diffraction }
\begin{figure*}[t]
    \centering
    \includegraphics[width=0.925\textwidth]{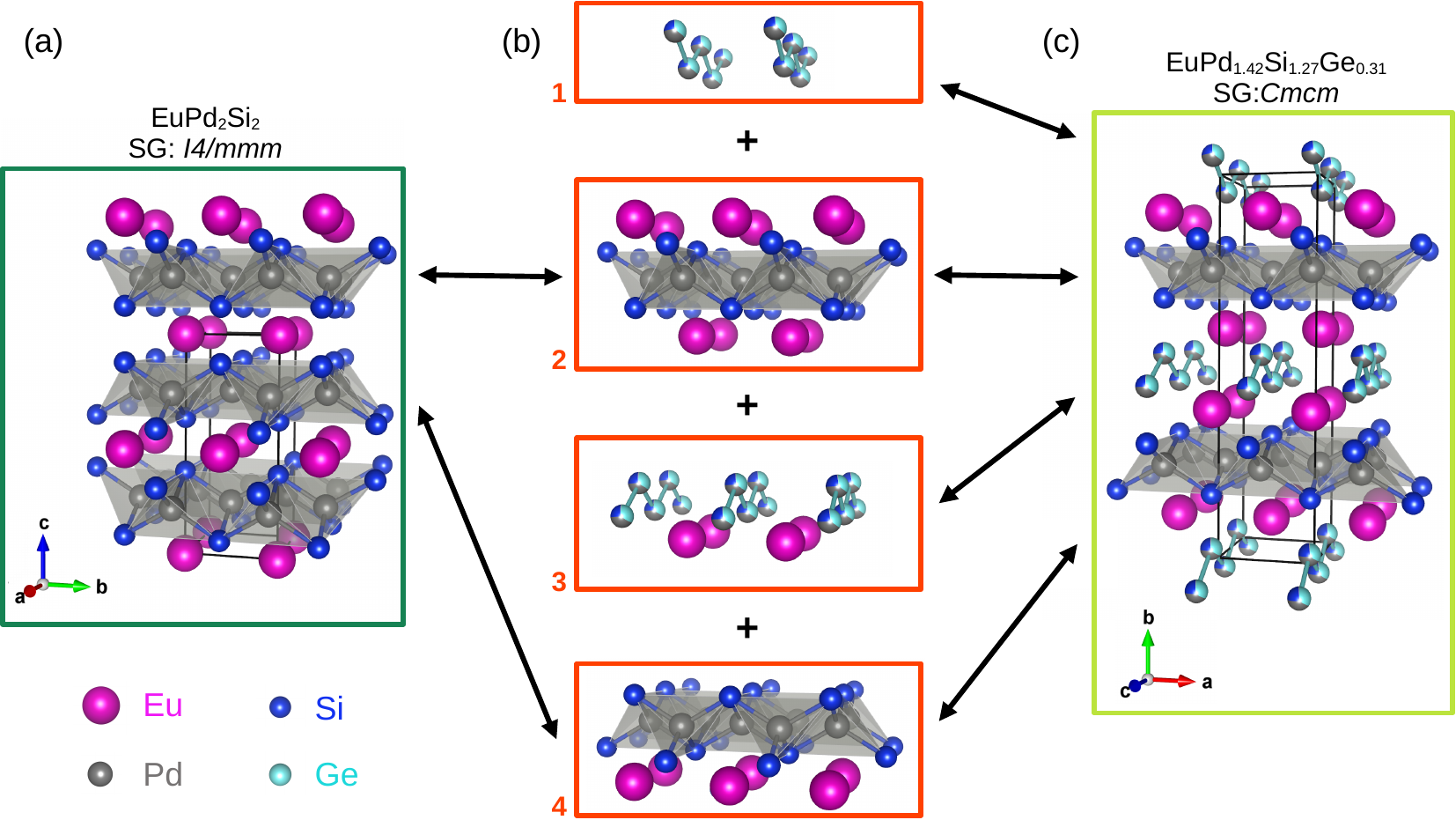}
    \caption{(a) Tetragonal unit cell of EuPd$_2$Si$_2$, (b) building blocks of EuPd$_2$Si$_2$ and of the newly discovered structure of the secondary phase, (c) orthorhombic unit cell of the EuPd$_{1.42}$Si$_{1.27}$Ge$_{0.31}$ secondary phase. As shown, the two structural blocks 2 and 4 are present in the unit cell of both EuPd$_2$Si$_2$ and EuPd$_{1.42}$Si$_{1.27}$Ge$_{0.31}$. Furthermore, the building blocks 1 and 3 are additionally present in the structure of EuPd$_{1.42}$Si$_{1.27}$Ge$_{0.31}$. Finally, the highly disordered (Si, Ge, Pd)-chain along the crystallographic $c$ direction repeats at the opposite end of the unit cell of EuPd$_{1.42}$Si$_{1.27}$Ge$_{0.31}$ as a consequence of the translational symmetry along $b$.} 
    \label{fig:XRD_structure}
\end{figure*}
The PXRD analysis of samples extracted from part A 
shows that in this part the ThCr$_2$Si$_2$ structure of the 122 target phase is maintained \cite{Peters2023}. In contrast, samples extracted from part B clearly show additional reflections from a side phase. We isolated a phase-pure sample of the side phase from part C, and its PXRD pattern (blue line) is shown in Fig.~\ref{fig:PXRD_MP805} in the supplemental material \cite{SI_Ausubst2025}. 
\subsubsection{Wavelength-dispersive x-ray spectroscopy}
According to the WDX analysis of a polished cut through the grown sample with $x_{\rm nom}=0.2$, Fig.~\ref{sample-MP809_WDX} in \cite{SI_Ausubst2025}, the length of part A ($l_{\rm A}^{0.2}=3.5\,\rm mm$) where the target phase crystallizes is strongly reduced when comparing it to a growth with $x=0.1$ ($l_{\rm A}^{0.1}=8\,\rm mm$, Fig.~1 in \cite{Peters2023}).
The elemental distribution of Ge determined using $\mu$-XRF spectroscopy shown in Fig.~\ref{elements}(c)
indicates the formation of Ge-rich side phase inclusions after $3.5\,\rm mm$. 
Through WDX analysis, we found a small change in the  stoichiometry from Eu\,:\,Pd\,:\,Si\,:\,Ge= 20.7\,:\,41.1\,:\,35\,:\,3.2 (next to the seed) to Eu\,:\,Pd\,:\,Si\,:\,Ge= 20.7\,:\,42.3\,:\,33.6\,:\,3.4 ($\approx 3$\,mm below the seed) of the Ge-substituted 122 compound in part A indicating a slight change in the Si-Pd ratio. A similar homogeneity range was observed for the unsubstituted system EuPd$_2$Si$_2$ before \cite{Kliemt2022a}.
The limited resolution of the WDX analysis does not allow to resolve the minor variations in the Ge content that are observed via $\mu$-XRF spectroscopy, Fig.~\ref{elements}.

\subsubsection{Magnetic susceptibility}
We studied the magnetic properties of the samples extracted from part A using VSM and found that these clearly show a valence crossover at $T_v^{\prime}=54\,\rm K$, Fig.~\ref{MP809_MvT_x0p2_Ge-subst_EuPd2Si2_Tv} in \cite{SI_Ausubst2025}. However, other methods such as HAXPES \cite{Fedchenko2024a} found no indication of a valence crossover but instead magnetic order for the same samples. 
The minor variations of the Ge content are resolved through $\mu$-XRF, Fig.~\ref{elements}(c), and are probably responsible for the different behavior of samples extracted from the same large Czochralski-grown sample. 

\subsection{Antiferromagnetic order in EuPd$_{1.42}$Si$_{1.27}$Ge$_{0.31}$}

\subsubsection{Chemical analysis}
EDX and WDX analyzes were performed on various samples extracted from part C of grown EuPd$_2$(Si$_{1-x}$Ge$_x$)$_2$ ingots with $x_{\rm nom}=0.2$.
The WDX analysis yields a composition of the side phase of Eu : Pd : Si : Ge $\approx$ 25.8 : 37.6 : 28.5 : 7.5, Fig.~\ref{sample-MP809_WDX} in \cite{SI_Ausubst2025}, which is in good agreement with the EDX result of Eu : Pd : Si : Ge = 26 : 38 : 27 : 8.

\subsubsection{Single-crystal x-ray diffraction on EuPd$_{1.42}$Si$_{1.27}$Ge$_{0.31}$} \begin{table}[t]
 \caption{Crystallographic data of EuPd$_{1.42}$Si$_{1.27}$Ge$_{0.31}$ derived from single-crystal and powder data measured at 300\,K. The goodness of fit, GOF, and the corresponding reliability factor $wR_2/R_{wp}$, $R_1/R_p$ are given as well.}
  \label{XRDtable1}
  \begin{tabular}{cc}
    \hline\hline
    Parameters & SCXRD  \\
    \hline
    Chemical formula   &  EuPd$_{1.42}$Si$_{1.27}$Ge$_{0.31}$ \\
    X-ray density (g/cm$^3$) & 7.6473    \\
    Space group & $Cmcm$   \\
    Formula units, $Z$ &  4  \\
    $a$ ({\AA})  & 4.33256(6) \\
    $b$ ({\AA}) & 16.88891(18) \\
    $c$ ({\AA}) & 4.28930(5)  \\
    $\alpha$ & 90$^{\circ}$  \\
    $\beta$  & 90$^{\circ}$ \\
   $\gamma$ & 90$^{\circ}$ \\
   $V$(\AA$^3$) & 313.858(7) \\ \hline
   GOF & 1.25   \\
   $wR_2/R_{wp}$ (\%) &  4.19  \\
   $R_1/R_p$ (\%) & 1.61   \\
    \hline\hline
  \end{tabular}
\end{table}\noindent
We used a small single-crystalline piece with approximate dimensions of 60\,$\times$\,40\,$\times$\,50\,$\mu$m$^3$ extracted from part C of a large grown ingot similar to the one shown in Fig.~\ref{sample-MP809_Czochralski} to determine the crystal structure of the side phase using SC-XRD. In Fig.~\ref{fig:XRD_structure}(a), the tetragonal unit cell with the space group $I4/mmm$ of EuPd$_2$Si$_2$ is illustrated, along with some building blocks in (b) that we will discuss below, and the newly discovered structure of the side phase, EuPd$_{1.42}$Si$_{1.27}$Ge$_{0.31}$, in (c). For the latter, the symmetry is reduced to the orthorhombic space group $Cmcm$ with lattice parameters $a$ = 4.33256(6)\,\AA, $b$ = 16.88891(18)\,\AA, $c$ = 4.28930(5)\,\AA, $\alpha$ = $\beta$ = $\gamma$ = 90$^{\circ}$ and a volume of 313.858(7) \AA$^3$. As can be quickly seen, the unit cell of EuPd$_{1.42}$Si$_{1.27}$Ge$_{0.31}$ is composed of various building blocks, which are shown individually in Fig.~\ref{fig:XRD_structure}(b). Upon closer inspection of Figs.~\ref{fig:XRD_structure}(a), (b), and (c), it can be seen that the unit cells of EuPd$_2$Si$_2$ and EuPd$_{1.42}$Si$_{1.27}$Ge$_{0.31}$ are related to each other: 
The unit cell of EuPd$_2$Si$_2$ can be decomposed into the two-dimensional network of PdSi$_4$ tetrahedra with the respective Eu layers above and below (building block 2 in Fig.~\ref{fig:XRD_structure}(b)), as well as the mirror-symmetric network of PdSi$_4$ tetrahedra (related by the mirror plane through the lower Eu layer of building block 2), with the corresponding Eu layer below (building block 4 in Fig.~\ref{fig:XRD_structure}(b)). 
Both structural blocks also exist in EuPd$_{1.42}$Si$_{1.27}$Ge$_{0.31}$, however, here a zig-zag chain (building block 1 in Fig.~\ref{fig:XRD_structure}(b)), occupied by a high degree of disorder of 42 \% Pd, 27 \% Si, and 31 \% Ge atoms (derived from our refinement), is located above building block 2, and another such disordered chain (building block 3 in Fig.~\ref{fig:XRD_structure}(b)), along with an additional layer of Eu atoms, is situated between building blocks 2 and 4. Finally, the unit cell terminates with a translation-symmetric disordered zig-zag chain of Pd, Si, and Ge atoms (in other words, the repetition of building block 1). 
As the disorder is confined to the zigzag chains, one might describe the composition of the compound as EuPdSi$X$ with $X$=Pd$_{0.42}$Si$_{0.27}$Ge$_{0.31}$. 
The tetragonal symmetry in EuPd$_2$Si$_2$ is primarily broken in orthorhombic EuPd$_{1.42}$Si$_{1.27}$Ge$_{0.31}$ by the insertion of these zig-zag chains running along the crystallographic $c$ direction, thereby leading to nonequivalent $a$ and $c$ directions. The refined crystallographic results obtained from our SC-XRD measurements on the secondary phase EuPd$_{1.42}$Si$_{1.27}$Ge$_{0.31}$ are summarized in Tables \ref{XRDtable1} and \ref{XRDtable2}. 
\begin{table}[t]
\caption{Atomic species, Wyckoff positions, coordinates, occupation numbers, Occ., and equivalent atomic displacement parameters (ADPs), $U_{\mathrm{eq}}$, of EuPd$_{1.42}$Si$_{1.27}$Ge$_{0.31}$ derived from single-crystal XRD measured at 300 K. The ADPs were refined anisotropically but due to space limitations only the $U_{\rm eq}$ are listed in the Table. During the refinement, the coordinates and the ADPs of the disordered chain atoms Ge(1), Pd(2), and Si(2) were constrained together, and the sum of the occupancies of these atoms was fixed to 1.} 
  \centering
  \label{XRDtable2}
    \begin{tabular}{ccccccc}
    \hline\hline
    Atom  & Wyckoff & $x$ & $y$ & $z$ & Occ. & $U_{\mathrm{eq}}$ ({\AA}$^{2}$)  \\
    \hline
    Eu(1) & 4$c$ & 0 & 0.39765(1) & 1/4 & 1 &  0.00914(2)\\
    Pd(1) & 4$c$ & 1/2 & 0.74964(1) & 3/4 & 1 & 0.01135(2)   \\
    Si(1) & 4$c$ & 1/2 & 0.68013(3) & 1/4 & 1 & 0.00926(8)   \\
    Ge(1) & 4$c$ & 1/2 &  0.53846(1) & 1/4 &  0.309(2) & 0.01070(3)  \\
    Pd(2) & 4$c$ & 1/2 &  0.53846(1) & 1/4 &  0.424(2) & 0.01070(3)  \\
    Si(2) & 4$c$ & 1/2 &  0.53846(1) & 1/4 &  0.267(2) & 0.01070(3)  \\
    \hline\hline 
  \end{tabular} 
\end{table} \newline \noindent
Since both the valence transition and the magnetic interactions are predominantly governed by the bond lengths and atomic volumes of Eu, a detailed analysis of the interatomic distances is essential: EuPd$_{1.42}$Si$_{1.27}$Ge$_{0.31}$ has in-plane Eu-Eu distances of
4.3326 and 4.2893\,{\AA} along the $a$ and $c$ directions, respectively. These values are significantly higher than those of related valence crossover compounds, such as EuPd$_2$Si$_2$ (4.2402\,{\AA}). However, they follow a similar trend to that observed in EuPd$_2$(Si$_{1-x}$Ge$_x$)$_2$ where the Eu-Eu distance increases with higher Ge concentrations (for instance, to 4.2828\,{\AA} when $x$ = 0.17) \cite{Peters2023}. The latter compound, while not exhibiting valence transitions, displays AFM order at lower temperature. Similarly, the Eu-Eu distances across the PdSi$_4$ tetrahedron networks, which are stacked out-of-plane, are significantly larger at 5.8460\,{\AA} compared to those in EuPd$_2$Si$_2$ (5.7795\,{\AA}) and EuPd$_2$(Si$_{1-x}$Ge$_x$)$_2$ (5.8050\,{\AA} for $x$ = 0.17). Additionally, there is a shorter Eu-Eu distance of 4.0686\,{\AA} connecting two consecutive Eu layers across the zig-zag chains in EuPd$_{1.42}$Si$_{1.27}$Ge$_{0.31}$, a feature absent in EuPd$_2$Si$_2$ and EuPd$_2$(Si$_{1-x}$Ge$_x$)$_2$ (see Fig.~\ref{fig:XRD_structure}(a), (b) and (c)).\\
Finally, we carefully examine the volume associated with the Eu atom, $V_{\rm Eu}$, defined from the bond lengths to its nearest-neighbor Si, Ge and Pd atoms in the first coordination shell, occupying either tetrahedral (Si, Ge) or zigzag-chain (Si, Ge, Pd) positions. For EuPd$_2$Si$_2$ and EuPd$_2$(Si$_{1-x}$Ge$_x$)$_2$ ($x$ = 0.17), this Eu-centered polyhedral volume amounts to 43.51 and 44.38\,{\AA}$^3$, respectively. For EuPd$_{1.42}$Si$_{1.27}$Ge$_{0.31}$, the corresponding volume is considerably larger, $V_{\rm Eu}$ = 
56.55\,{\AA}$^3$. In addition, the incorporation of the zig-zag chain structural block not only breaks the tetragonal symmetry but also effectively constrains this volume. Considering that a trivalent Eu atom requires a markedly reduced coordination volume relative to a divalent Eu atom, the comparatively large and expectedly clamped $V_{\rm Eu}$ value observed for EuPd$_{1.42}$Si$_{1.27}$Ge$_{0.31}$ suggests that a valence change is highly improbable, but rather supports the presence of divalent magnetic Eu as observed in EuPd$_2$(Si$_{1-x}$Ge$_x$)$_2$ for higher Ge contents. \\
PXRD data of a powdered sample, which was also extracted from part C, are discussed in Ref. \cite{SI_Ausubst2025} (Fig.~\ref{fig:PXRD_MP805}). The data are in good agreement with the structure determined from SC-XRD.

\subsubsection{Magnetic properties of EuPd$_{1.42}$Si$_{1.27}$Ge$_{0.31}$}
\begin{figure}
\centering
\includegraphics[width=0.47\textwidth]{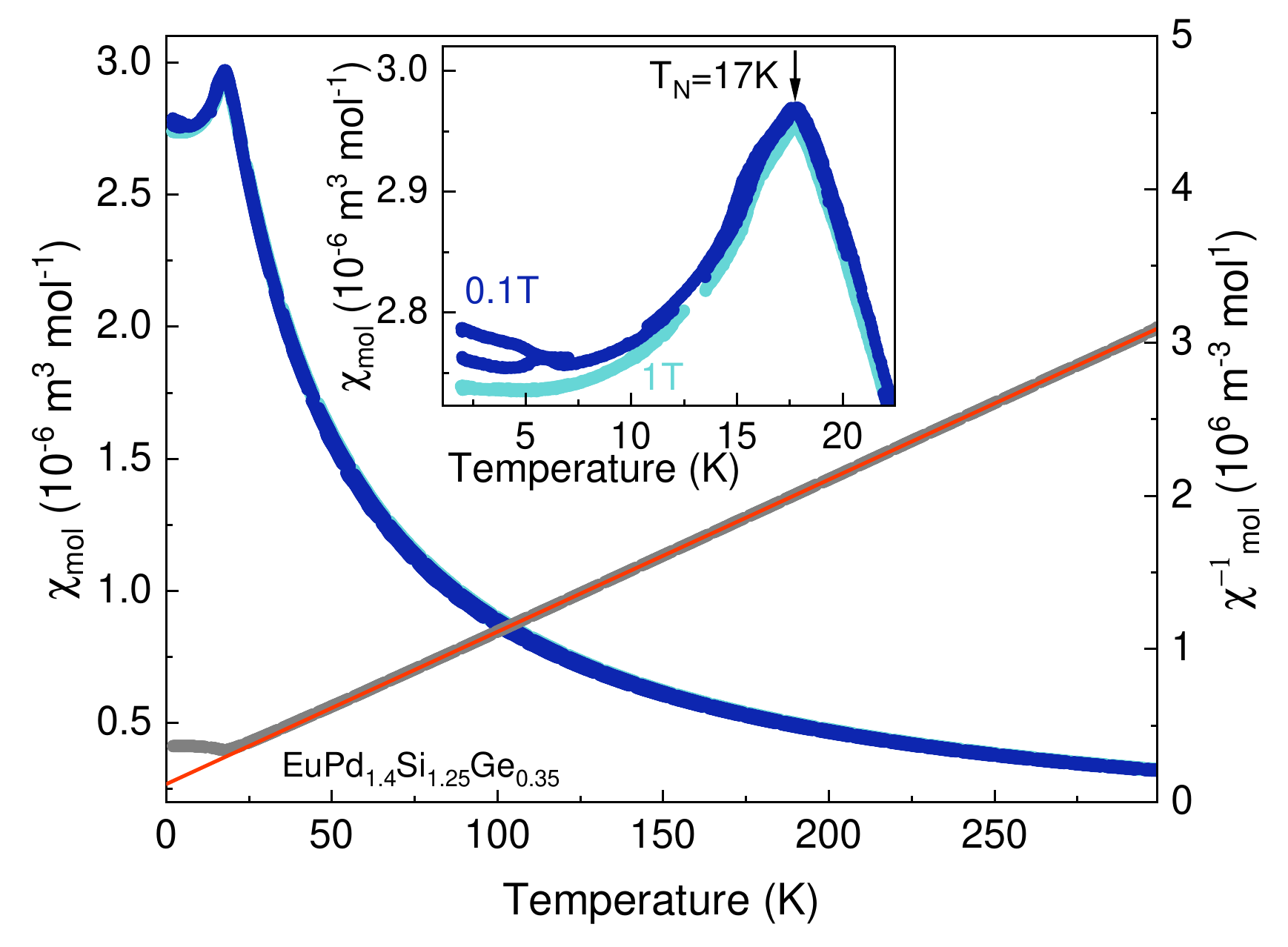}
	\caption[]{EuPd$_{1.42}$Si$_{1.27}$Ge$_{0.31}$: Temperature dependence of the magnetic susceptibility of a polycrystalline sample (left axis). The transition at T$_N$=17\,K is shown enlarged in the inset, with the measurement at 0.1 T displayed in dark blue and that at 1 T in light blue. The inverse susceptibility at $\mu_0H = 1$\,T is plotted on the right axis and the Curie-Weiss fit is shown in orange. }
\label{MvT_x0p2_Ge-subst_EuPd2Si2_NF}
\end{figure}\noindent
The inverse susceptibility (gray symbols) at $\mu_0H = 1$\,T is plotted in Fig.~\ref{MvT_x0p2_Ge-subst_EuPd2Si2_NF} on the right axis. The material shows Curie-Weiss behavior at high temperatures, and the solid red line represents a linear fit to the data in the temperature range $150\,{\rm K} < T < 300\,\rm K$. From the fit, a Weiss temperature of $\theta_W = (-12.0 \pm 0.5)$\,K 
and an effective magnetic moment of $\mu_{\rm eff}^{\rm exp} = (8.00 \pm 0.01)\,\mu_{\rm B}$ close to the calculated value of $\mu_{\rm eff}^{\rm calc}=7.94\,\mu_{\rm B}$ were determined. Comparison with previous magnetic susceptibility data \cite{Peters2023} shows that this side phase with its transition at $T_{\rm N}=17\,\rm K$ has appeared in the previous study.\\
The temperature dependence of the magnetic susceptibility of polycrystalline EuPd$_{1.42}$Si$_{1.27}$Ge$_{0.31}$ is shown in Fig.~\ref{MvT_x0p2_Ge-subst_EuPd2Si2_NF} (left axis) for a field of $\mu_0H = 0.1$\,T (light blue symbols) and 1\,T (dark blue symbols). 
The material undergoes an antiferromagnetic transition at $T_{\rm N}=17\,\rm K$. This ordering is only weakly affected by the magnetic fields investigated here. At low fields, a transition caused by an additional impurity phase is observed, which has also been discussed earlier \cite{Peters2023}.

\section{Discussion}
In Eu(Pd$_{1-x}$Au$_x$)$_2$Si$_2$, Au is incorporated at the Pd site \cite{Gupta1982} and this substitution is not isoelectronic.
This contrasts to our previous study on the series EuPd$_2$(Si$_{1-x}$Ge$_x$)$_2$ \cite{Peters2023} where Si is partially replaced by isoelectronic Ge. The effect of Au substitution on the band structure revealed the formation of additional bands due to the non-isoelectronic substitution \cite{Fedchenko2024a}. 
For the crystal growth of Au-substituted EuPd$_2$Si$_2$ which forms incongruently, we used the Czochralski method which has the advantage, that the crystal can be separated from the flux directly during the growth. The chemical analysis of the samples clearly shows that under the experimental conditions chosen here, the existence region of the 122 phase substituted with Au is limited. In future work, further modifications of the initial melt stoichiometry are required to increase the level of Au substitution and tune the system to the possible critical endpoint. \\
We found that the rate of incorporation of Au in the structure is lower than that of Ge \cite{Peters2023} and yields an enrichment of Au in the melt which leads to the formation of different side phases in the course of the growth. Already for $x_{\rm nom}=0.2$, which corresponds to $x_{\rm WDX}=0.065\pm 0.005$ the side phase fraction in the growth is dominant. This is in contrast to the series of isoelectronic substitutions with Ge. Here, $x_{\rm nom}=0.30$ samples with a much higher Ge content of $x_{\rm WDX}=0.154\pm0.009$ could be grown where the 122 structure was maintained \cite{Peters2023}. In the case of Ge substitution, samples in the 122 structure with high Ge content $x_{\rm WDX}=0.105\pm0.008$ order antiferromagnetically below $T_N=47\,\rm K$ \cite{Peters2023}. This is different in the case of the Au substitution presented here where the highest substituted samples still show a valence crossover.

Our finding cannot be directly compared to a result presented in Ref.~\cite{Croft1982, Segre1982} where AFM order was detected for a higher Au substitution level of $x_{\rm nom}=0.2$. The annealed polycrystalline samples studied there showed a first-order transition for $x_{\rm nom}=0.1, 0.125$ and 0.15 but were prepared differently and it is reasonable to assume that there $x_{\rm nom}$ was probably equal to $x_{\rm real}$ although no results of a chemical analysis were given to determine the real amount of Au incorporation.
The properties of pure as well as substituted EuPd$_2$Si$_2$ are very sensitive to slight changes in sample stoichiometry, so that samples prepared under different conditions are not comparable between the different studies.
In the Au-substitution series, the lowest crossover temperature $T^{\prime}_v=77\,\rm K$ is found for $x_{\rm nom}=0.15$. 
This is higher than for the Ge substitution, where a valence crossover is found down to $T^{\prime}_v=54\,\rm K$. Comparing the valence crossover temperatures with those determined in~\cite{Segre1982}, we assume that in our single crystals the real substitution level is lower compared to polycrystalline Au substituted samples with $x^{\rm poly}=0.15$ since a much lower $T^{\prime}_v=39\,\rm K$ was reached there. Our substitution level is probably similar to $x^{\rm poly}=0.125$ with $T^{\prime}_v\approx 64\,\rm K$ presented in \cite{Gupta1982}.
For $x_{\rm nom}=0.2$, the HC peak  occurs unexpectedly at higher $T$ than for the $x_{\rm nom}=0.15$ and no signs of magnetic order of the target phase are observed. Instead, at least two distinct transitions appear which indicate magnetic order of side phases at $T^1_m=11\,\rm K$ and $T^2_m=50\,\rm K$. 
Our investigation of samples of EuPd$_2$(Si$_{1-x}$Ge$_x$)$_2$ with $x_{\rm nom}=0.2$ revealed that there are minor changes in Ge concentration in samples that are at the edge of the valence crossover close to showing magnetic order. $\mu$-XRF has turned out to be the method of choice when it comes to resolving these minor variations which were not detectable by using other methods like EDX or WDX.  

\section{Summary}

We present the details of the crystal growth experiments of  Eu(Pd$_{1-x}$Au$_x$)$_2$Si$_2$ using the Czochralski method from a levitating melt and showed the results of WDX and PXRD analysis for different $x$ which reveal a decreasing region of existence of the 122 phase upon increasing Au content in the initial melt. For the experimental conditions used here, at maximum an Au content of $x_{\rm WDX}=0.065\pm 0.005$ can be achieved in the samples. For higher Au content in the melt, the formation of Eu-rich side phases occurs, which strongly influence the magnetic susceptibility.
For $x_{\rm nom}=0.1$, the growth of a large ($\approx$10\,mm) sample without or with low side phase content was possible, and the crossover temperatures determined by LT-PXRD, HC and VSM perfectly match.
For higher Au content, the fraction of the 122 phase that was pulled from the melt was reduced until only a small sample ($\approx$1\,mm) of the target phase could be grown for $x_{\rm nom}=0.2$ due to limited Au incorporation.
In accordance with the literature, for increasing Au content in the samples an increase of the $a$ and $c$ lattice parameters \cite{Abd-Elmeguid1985_ZfPB} and a shift of $T^{\prime}_v$ towards lower temperatures \cite{Gupta1982, Croft1982, Segre1982} is observed in the magnetic susceptibility and heat capacity data. All samples studied show a valence crossover transition. Signatures of a first-order transition were not observed for any sample in the system. 

In contrast to published data up to a nominal Au concentration $x_{\rm nom}=0.2$ \cite{Croft1982, Segre1982}, we do not observe a transition into an AFM phase. Furthermore, we cannot confirm the existence of a first-order valence transition as reported for $x_{\rm nom}=0.1, 0.125$ and 0.15 in Refs.~\cite{Gupta1982, Croft1982, Segre1982}.
We cannot exclude the possibility that these differences appear due to the different procedures of preparation. 
We showed that using the Czochralski method large Eu(Pd$_{1-x}$Au$_x$)$_2$Si$_2$ single crystals with $x_{\rm nom}=0.1$ with a low side phase fraction can be grown. Future work will be dedicated to adjust the experimental conditions to further increase the Au substitution level in order to grow samples closer to the expected critical endpoint.  

Several samples of EuPd$_2$(Si$_{1-x}$Ge$_x$)$_2$ with $x_{\rm nom}=0.2$
were grown by the Czochralski method and analyzed in detail with respect to their chemical and structural properties. Samples with this Ge substitution level are at the edge of the valence crossover regime and are potential candidates for the study of critical elasticity. In previous studies \cite{Peters2023}, it turned out that such samples close to the valence-crossover to magnetism transition often show double transitions in their physical properties. Here we identified minor variations in the Ge concentration detected via $\mu$-XRF as a possible origin.
Furthermore, it was possible to identify EuPd$_{1.42}$Si$_{1.27}$Ge$_{0.31}$ as the side phase that appears in the grown samples. This quarternary compound shows AFM order at $T_{\rm N}=17\,\rm K$.

\section*{Data Availability}
The data sets are available via https://doi.org/xxxxxx at the open data repository of the Goethe University Frankfurt (GUDe) \cite{}. Structural data are available from the Karlsruhe Institute of Technology repository KITOpen via https://doi.org/ [zz].
\\
\begin{acknowledgments}
We thank Siegmar Roth and Andre Beck at the Institute for Quantum Materials and Technologies, Karlsruhe Institute of Technology and 
Tim F\"orster and Franz Ritter at Goethe university Frankfurt for their great technical support. We acknowledge funding by the Deutsche Forschungsgemeinschaft (DFG, German Research Foundation) via the TRR 288 (422213477, project A03, B03).   
\end{acknowledgments}

\section{References}
\bibliography{Bib}
\cleardoublepage
\appendix

\renewcommand{\thetable}{S\Roman{table}}\renewcommand{\thefigure}{S\arabic{figure}}
\setcounter{figure}{0}
\begin{center}\textbf{Supplemental Information \\}\end{center}

\section{Au-substituted EuPd$_2$Si$_2$} 
\subsection{Parameters of the crystal growth} \noindent
Tab.~\ref{tab:3} summarized the Au-substitution experiments carried out with various nominal concentrations x$_{\rm nom}$ and the chosen stoichiometries. The solidus and liquidus temperatures were determined pyrometrically from observations made during the crystal growth process. The values for the unsubstituted compound are taken from earlier differential thermal analysis (DTA) measurements reported in Ref.~\cite{Kliemt2022a}.
\begin{table}[htbp]
\begin{center}
\begin{tabular}{|c|c|c|c|c|c|}
\hline
$x_{\rm nom}$&Stoichiometry&$T_{\rm sol}$ &$T_{\rm liq}$\\
   & $\rm Eu : Pd : Au : Si $ & $[^\circ\rm C]$ &  $[^\circ\rm C]$          \\
\hline
0  &$1.45: 2.0: 0.0: 2.0$&1186 (DTA) \cite{Kliemt2022a}&1165 (DTA)\cite{Kliemt2022a}  \\
0.1  &$1.45: 1.8: 0.2: 2.0$&1160 $\pm$ 20&1210 $\pm$ 20\\
0.15 &$1.45: 1.7: 0.3: 2.0$&1120 $\pm$ 20&1180 $\pm$ 20\\
0.2    &$1.45: 1.6: 0.4: 2.0$&1150 $\pm$ 20&1200 $\pm$ 20\\
\hline
\end{tabular}
\end{center}
\caption{Eu(Pd$_{1-x}$Au$_x$)$_2$Si$_2$: Initial melt compositions and parameters of the crystal growth. }
\label{tab:3}
\end{table}
\subsection{Powder X-ray diffraction}
\begin{figure}[ht!]
    \centering
     \includegraphics[width=0.8\linewidth]{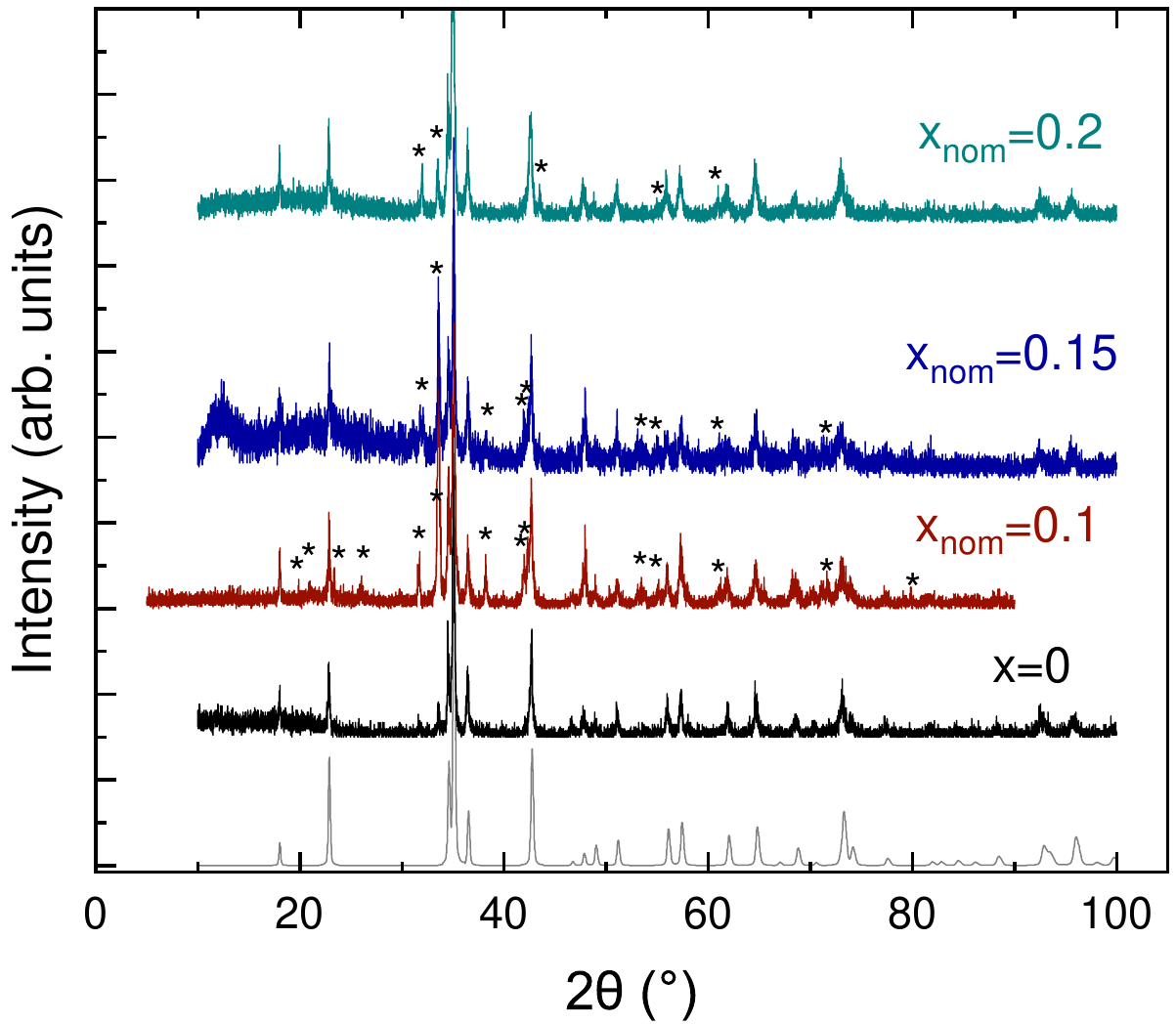}
	\caption[]{Room-temperature PXRD data of Eu(Pd$_{1-x}$Au$_x$)$_2$Si$_2$, $x=0$ data from \cite{Kliemt2022a} and with reference data for pure EuPd$_2$Si$_2$ (gray) \cite{Abd-Elmeguid1985_ZfPB}. Asterisks mark additional reflexes of the side phases.}
\label{PXRD_Au-subst_EuPd2Si2}
\end{figure}

\begin{figure}[ht!]
    \centering
     \includegraphics[width=0.9\linewidth]{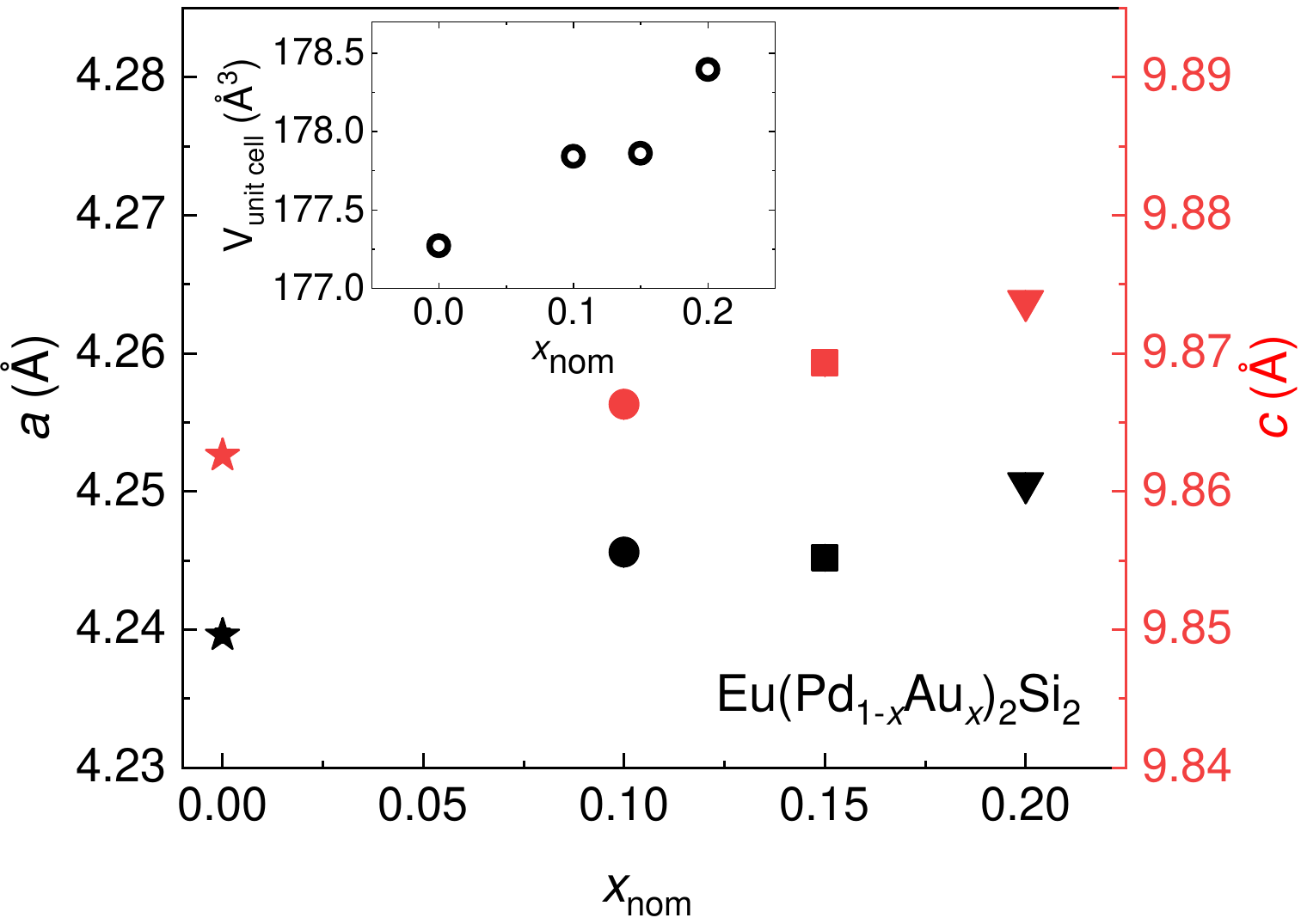}
	\caption[]{Room-temperature lattice parameters of Eu(Pd$_{1-x}$Au$_x$)$_2$Si$_2$: $a$ and $c$ increase with $x_{\rm nom}$. The lattice parameters were extracted from the PXRD data presented in  Fig.~\ref{PXRD_Au-subst_EuPd2Si2}. In the inset, the unit cell volume is shown with respect to $x_{\rm nom}$. $x=0$ data were taken from \cite{Kliemt2022a}.} 
\label{Lattice_Au-subst_EuPd2Si2}
\end{figure}

\begin{figure}[ht!]
    \centering
    \includegraphics[width=0.9\linewidth]{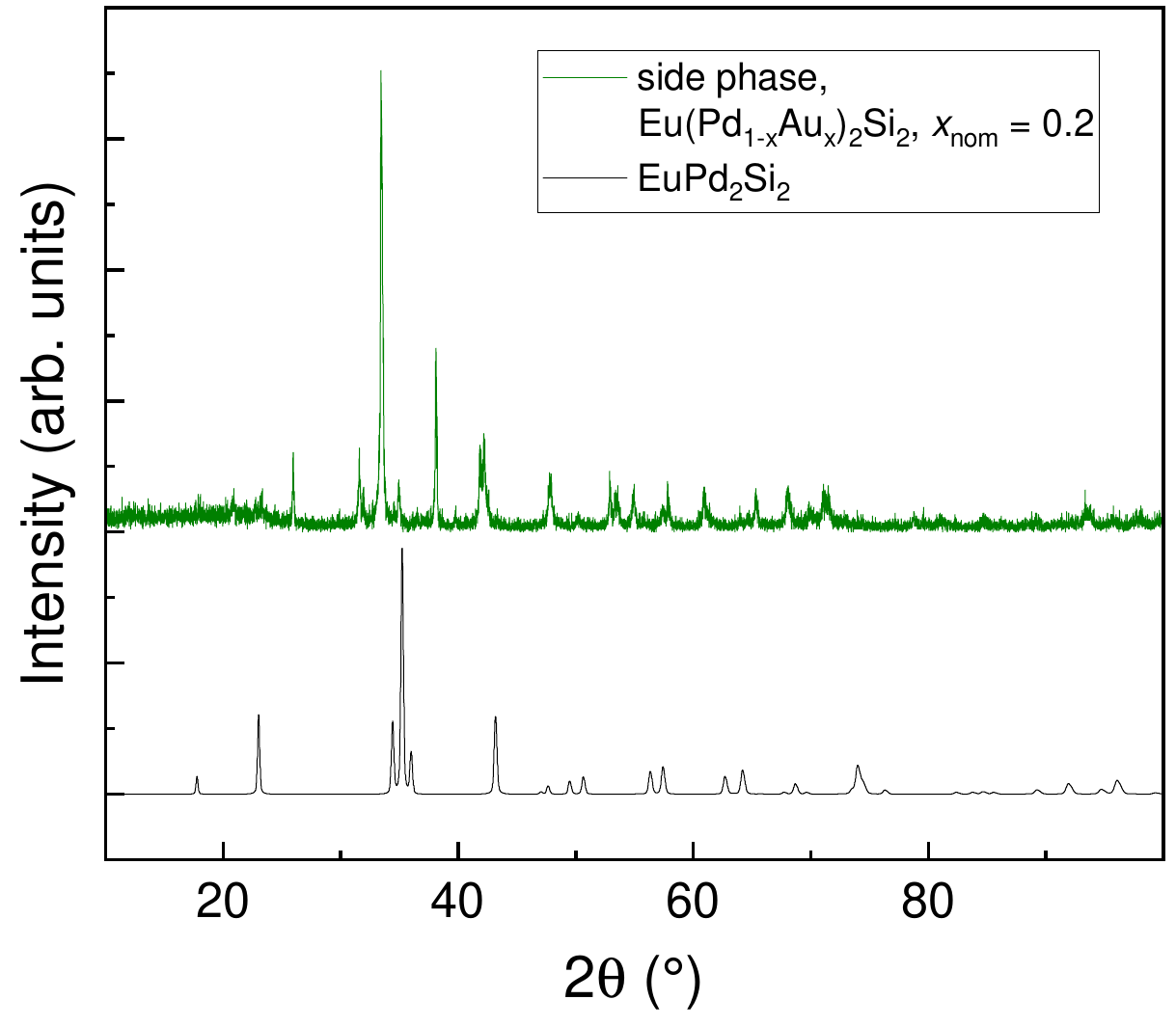}
\caption[]{Room-temperature PXRD data of the pure side phase in the Eu(Pd$_{1-x}$Au$_x$)$_2$Si$_2$, $x_{\rm nom}=0.2$ growth experiment in comparison with reference data for pure EuPd$_2$Si$_2$ \cite{Abd-Elmeguid1985_ZfPB}.  } 
\label{sidephase_0.2_RM102_PXRD}
\end{figure}

\begin{figure}[ht!]
    \centering
    \includegraphics[width=1.0\linewidth]{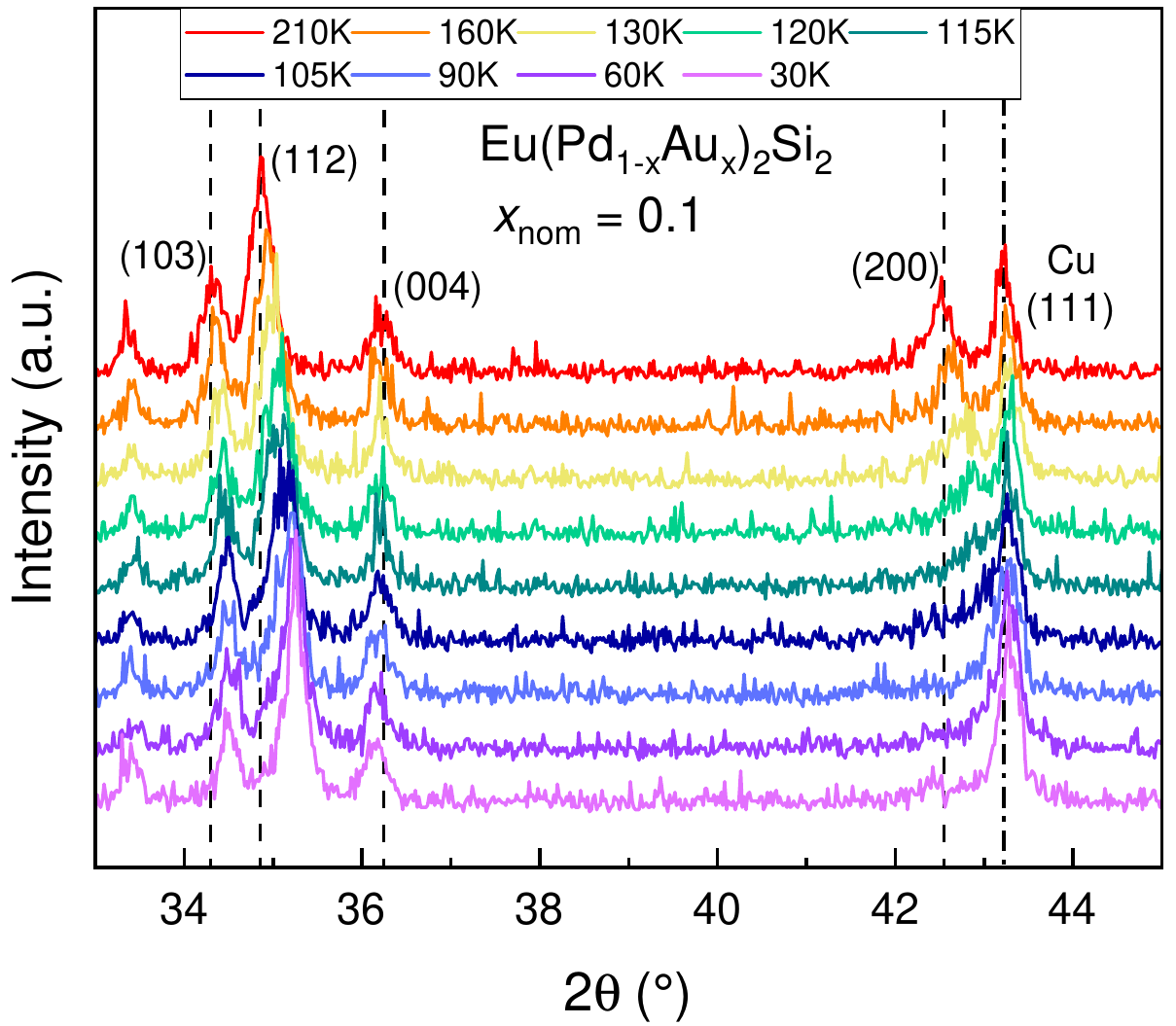}
\caption[]{Eu(Pd$_{1-x}$Au$_x$)$_2$Si$_2$,  $x_{\rm nom}=0.1$ low temperature PXRD. Lattice parameters are shown in Fig.~\ref{TTPXRD_Lattice_Au-subst_EuPd2Si2}. Additional Cu reflexes originate from the sample platform and can be used as a reference. }
\label{TT-PXRD_Au-subst_EuPd2Si2}
\end{figure}

\noindent
The PXRD data taken on different Au-substituted samples are shown in Fig.~\ref{PXRD_Au-subst_EuPd2Si2}. Additional reflexes that arise from side phases are marked by asterisks. The room-temperature lattice parameters for different substitution levels are shown in Fig.~\ref{Lattice_Au-subst_EuPd2Si2}. 
\noindent
The diffraction pattern of a pure sample of the side phase in the $x_{\rm nom}=0.2$ growth is shown in Fig.~\ref{sidephase_0.2_RM102_PXRD}. The reflexes of this side phase match most of the additional reflexes that are found in the $x_{\rm nom}=0.1$ and $x_{\rm nom}=0.15$ diffraction patterns in Fig.~\ref{PXRD_Au-subst_EuPd2Si2}.  \\
\newline \noindent
LT-PXRD data for $x_{\rm nom}=0.1$ are shown in Fig.~\ref{TT-PXRD_Au-subst_EuPd2Si2}. An additional reflex (Cu (111)) stems from the sample holder. While the position of the (004) reflex remains constant upon cooling, the (200) reflex strongly shifts to higher angles, indicating a contraction of the $a$ lattice parameter. The lattice parameters were determined by refinement and are shown in Fig.~\ref{TTPXRD_Lattice_Au-subst_EuPd2Si2} in the main text. 
\subsection{WDX and EDX} 
\begin{figure}[ht!]
    \centering
    \includegraphics[width=0.99\linewidth]{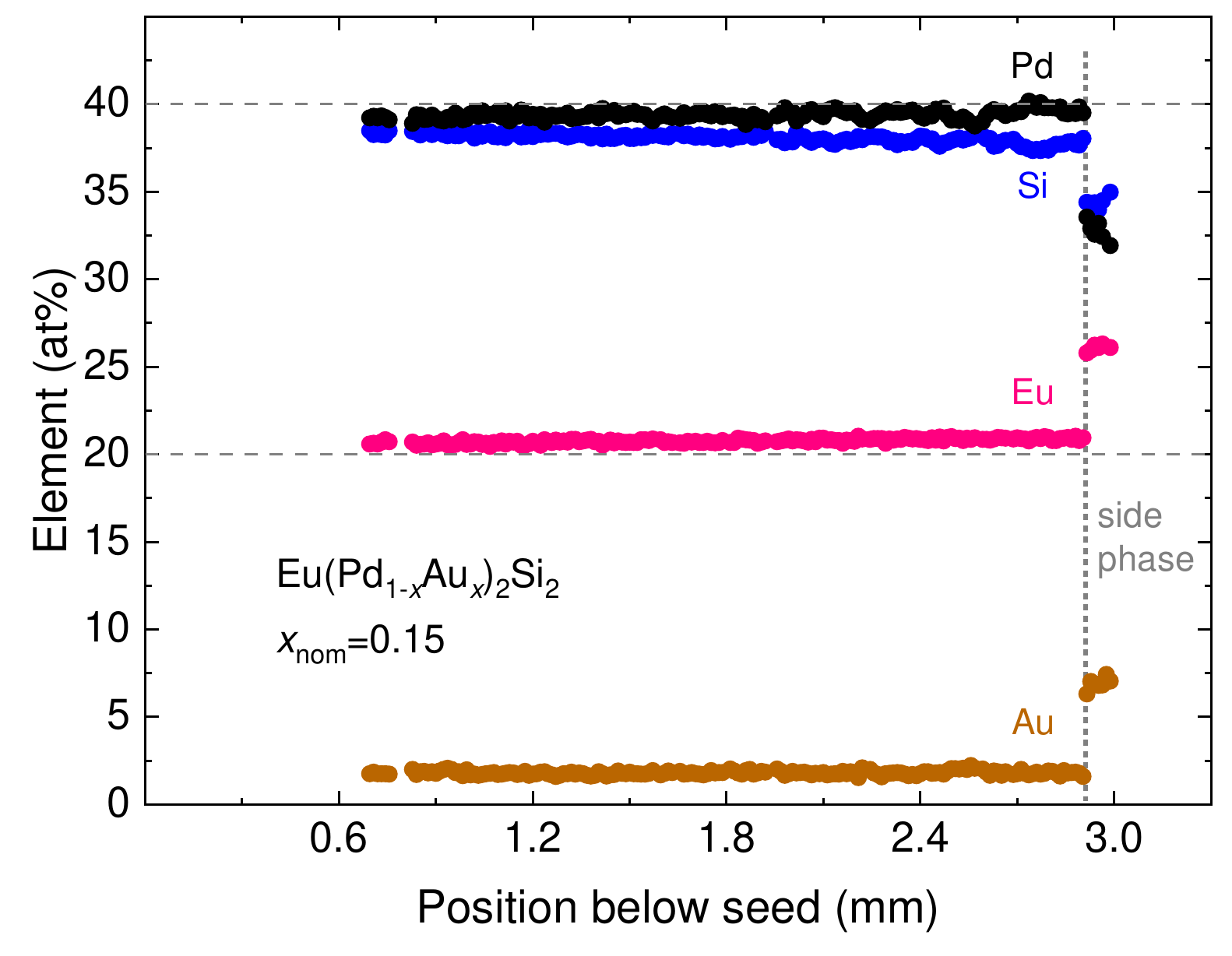}
\caption[]{Eu(Pd$_{1-x}$Au$_x$)$_2$Si$_2$: Results of the  WDX analysis of a polished longitudinal cut through a sample with $x_{\rm nom}=0.15$. Unreliable data points were removed for clarity.}
\label{WDX_0p15_Au-subst_EuPd2Si2}
\end{figure}

\noindent
The chemical composition of the Au-substituted samples determined by WDX data is summarized in  Tab.~\ref{tab:WDX}.
\begin{table*}[htbp]
\begin{center}
\begin{tabular}{|c|c|c|c|c|c|c|c|}
\hline
$x_{\rm nom}$	&Eu (WDX)&	Pd (WDX)&	Si (WDX)&	Au (WDX)&$x_{\rm WDX}$	&$x_{\rm EDX}$\\
      	&at\%	&	at\%	&	at\%	&	at\%	&&	\\	
\hline
0.1&	20.76$\pm$0.11&40.95$\pm0.13$&37.17$\pm$0.19&1.12$\pm$0.08	&0.028$\pm$0.004&0.040$\pm$0.020\\	
0.15&	20.62$\pm$0.10&39.45$\pm$0.19&37.8$\pm$0.22&	1.84$\pm$0.10&	0.046$\pm$0.005&0.043$\pm$0.013\\
0.2&	21.01$\pm$0.09&38.68$\pm$0.25&37.71$\pm$0.25&	2.6$\pm$0.11&	0.065$\pm$0.005&0.050$\pm$0.008\\
\hline
\end{tabular}
\end{center}
\caption{Eu(Pd$_{1-x}$Au$_x$)$_2$Si$_2$: WDX results of the target phase}
\label{tab:WDX}
\end{table*}
In Fig.~\ref{WDX_0p15_Au-subst_EuPd2Si2}, the WDX data taken on the polished surface of a longitudinal cut through a $x_{\rm nom}=0.15$ sample are shown. Throughout the sample, constant Eu and Au contents were detected. Furthermore, a slight shift of the Pd-Si composition along the sample, similar to the finding in the case of the pure compound \cite{Kliemt2022a}. After a length of $3\,\rm mm$, the growth of a side phase started.\\
In all experiments, side phases appeared during growth. The composition of the side phase was determined by WDX and is summarized in Tab.~\ref{tab:WDX_sidephase}.\\

\begin{table}[ht!]
    \centering
    \begin{center}
\begin{tabular}{|c|c|c|c|c|c|c|c|}
\hline
	$x_{\rm nom}$	&Eu (WDX)&	Pd (WDX)&	Si (WDX)&	Au (WDX)\\
      		&at\%	&	at\%	&	at\%	&	at\%		\\	
\hline
	0.1&	26.2$\pm$0.3&33.0$\pm0.4$&34.6$\pm$0.6&6.2$\pm$0.2\\	
0.15&	26.1$\pm$0.4&33.1$\pm$0.5&34.0$\pm$0.6&	6.8$\pm$0.2\\
	0.2&	26.1$\pm$0.5&29.8$\pm$0.7&34.1$\pm$0.9&	10.0$\pm$0.9\\
\hline
\end{tabular}
\end{center}
\caption{Side phases: WDX results}
\label{tab:WDX_sidephase}
\end{table}
\subsection{Magnetic susceptibility} 
\begin{figure}[ht!]
\centering
\includegraphics[width=0.9\linewidth]{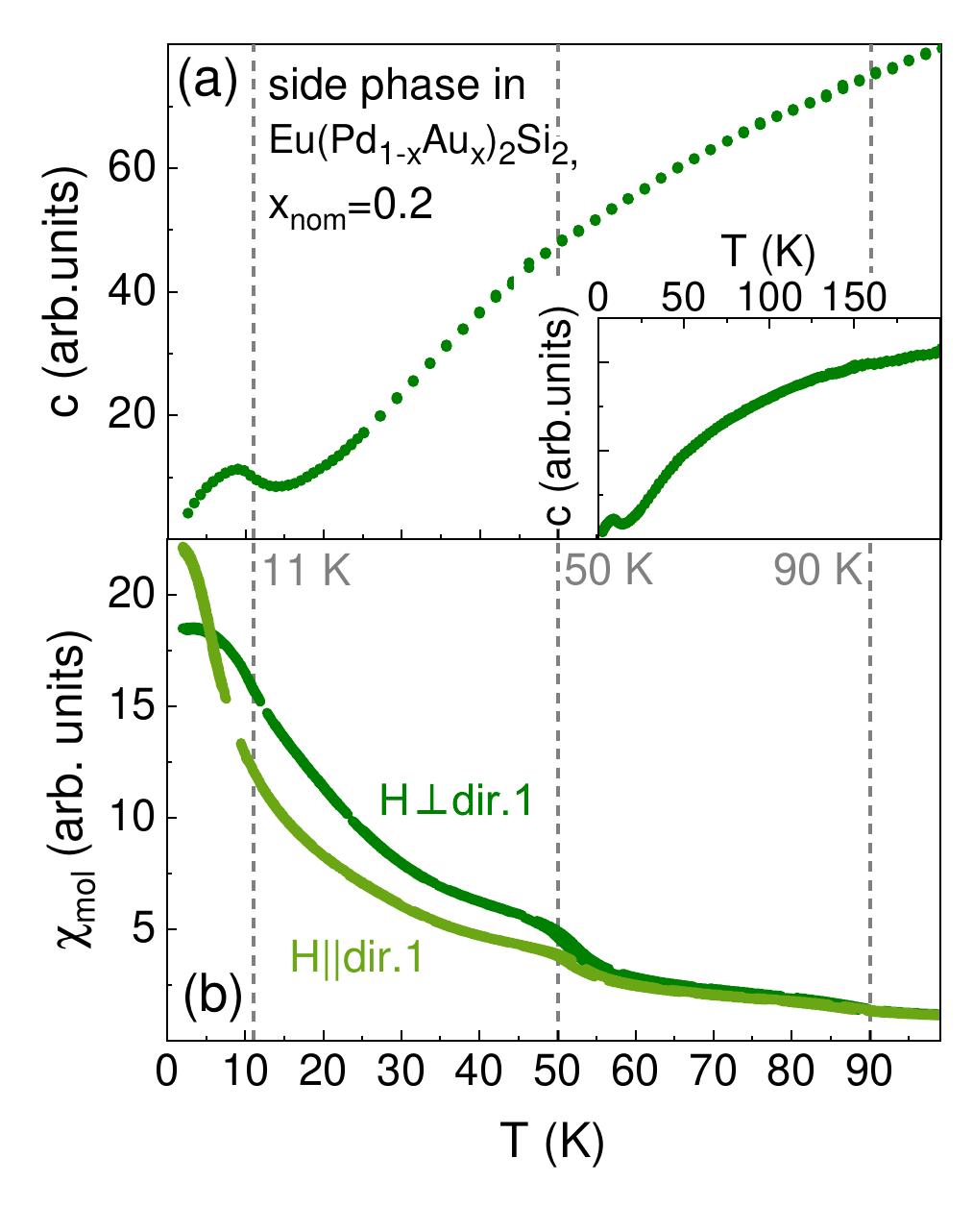}
	\caption[]{Heat capacity (a) and magnetic susceptibility measured with field applied along two perpendicular crystallographic directions (b) of a sample from the growth of Au-substituted EuPd$_2$Si$_2$, $x_{\rm nom}=0.2$ which consists mainly of side phase. }
\label{HC_Au-subst_EuPd2Si2_RM102b} 
\end{figure}

\noindent
The magnetic susceptibility of the side phase, Fig.~\ref{HC_Au-subst_EuPd2Si2_RM102b}(b) is anisotropic below 50\,K which indicates that the transitions at $T_m^1=11\,\rm K$, and $T_m^2=50\,\rm K$ are of magnetic origin.
\section{Ge-substituted EuPd$_2$Si$_2$}
\subsection{PXRD side phase}
\begin{figure}[ht!]
    \centering
    \includegraphics[width=0.99\linewidth]{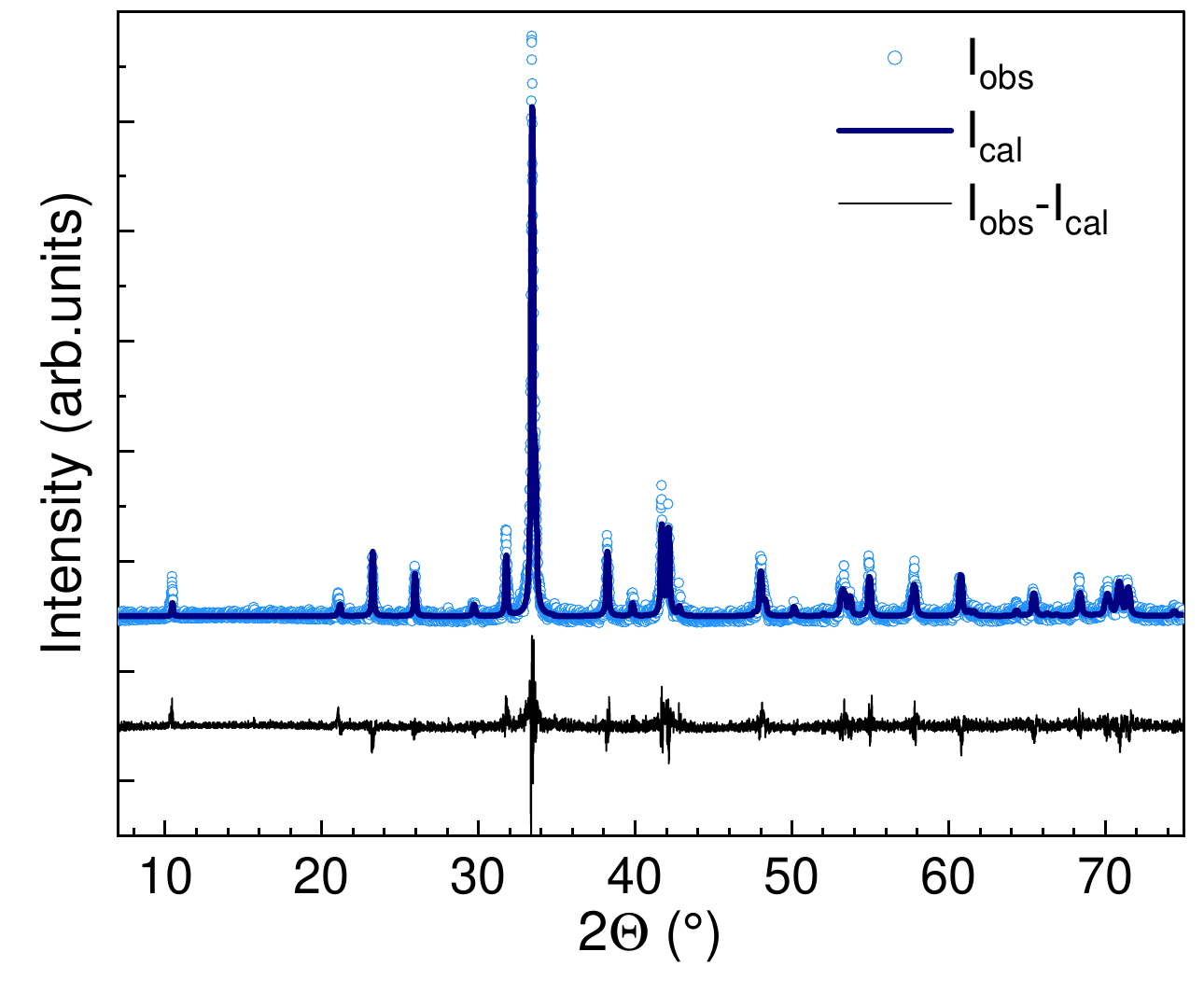}
    \caption{PXRD of crushed single crystals of section C from a growth of EuPd$_2$(Si$_{1-x}$Ge$_x$)$_2$ with $x_{\rm nom}=0.2$ (blue), the simulated data (black) according to the structure determined from SC XRD of EuPd$_{1.42}$Si$_{1.27}$Ge$_{0.31}$ is shown for comparison.}
    \label{fig:PXRD_MP805}
\end{figure}

\noindent
The PXRD data of the side phase formed in Ge-substituted EuPd$_2$Si$_2$ are shown in Fig.~\ref{fig:PXRD_MP805}. The measured diffraction pattern (light blue) is compared with the calculated pattern on the basis of the SC-XRD (dark blue). 
The obtained lattice parameters ($a$=4.3327\,\AA, $b$= 16.8914\,\AA, $c$= 4.2895\,\AA) show good agreement with those derived from SC-XRD.
\newline \noindent
\subsection{WDX and EDX}
\begin{figure}[ht!]
\centering
\includegraphics[width=0.44\textwidth]{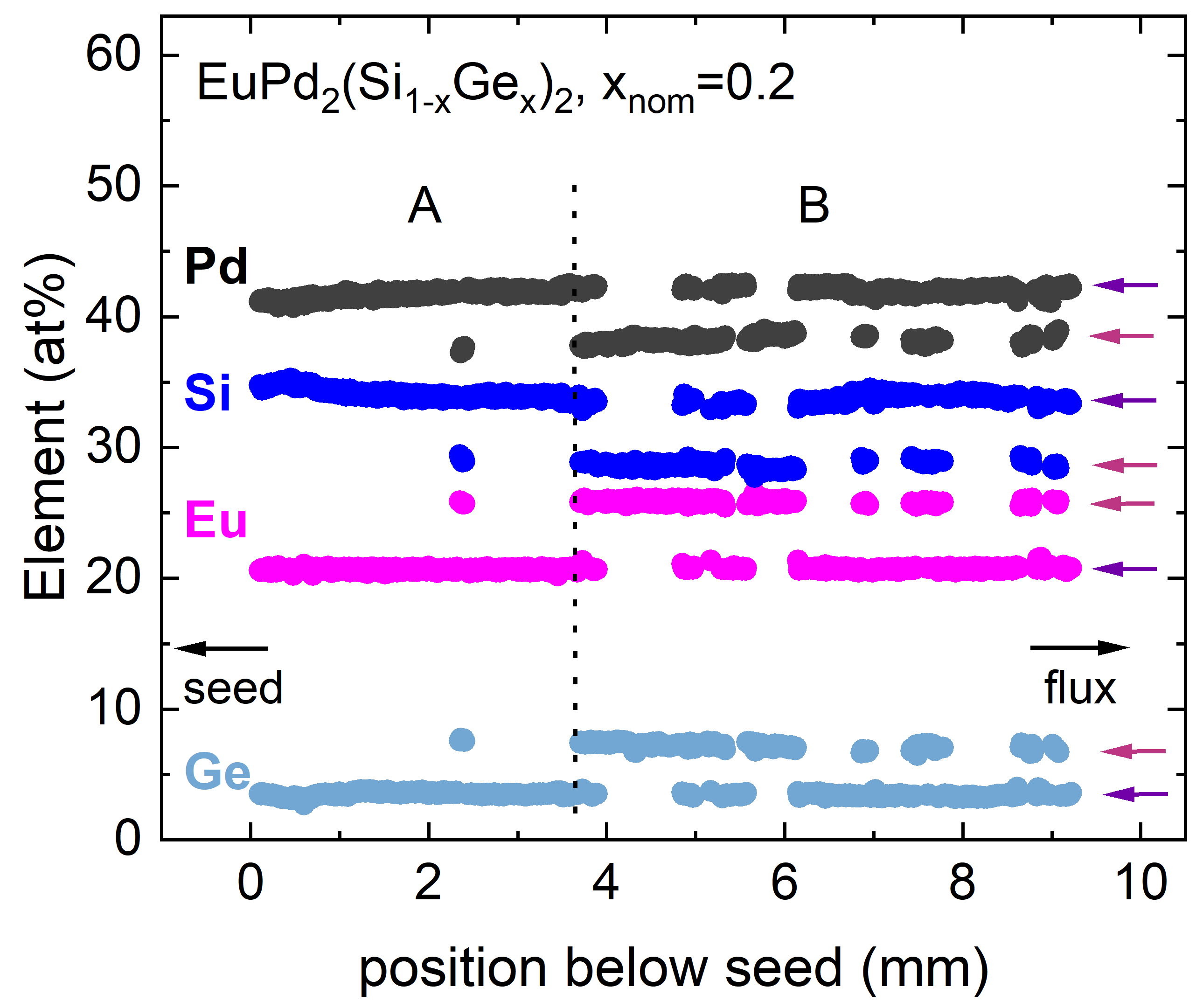}
	\caption[]{EuPd$_2$(Si$_{1-x}$Ge$_x$)$_2$, $x_{\rm nom}=0.2$: WDX analysis. The target-phase composition is marked by blue-violet arrows and side phase by red-violet arrows. }
\label{sample-MP809_WDX}
\end{figure}

WDX analysis of a longitudinal cut through a sample (polished surface)  EuPd$_2$(Si$_{1-x}$Ge$_x$)$_2$, $x_{\rm nom}=0.2$ grown by the Czochralski method is shown in Fig.~\ref{sample-MP809_WDX}
and confirms the formation of the target phase (blue-violet arrows) in part A and the formation of a side phase (red-violet arrows) in part B.
The WDX analysis reveals a composition of the side phase of Eu : Pd : Si : Ge $\approx$ 25.8 : 37.6 : 28.5 : 7.5, Fig.~\ref{sample-MP809_WDX} which is consistent with the EDX result of Eu : Pd : Si : Ge = 26 : 38 : 27 : 8.

\subsection{Magnetic susceptibility EuPd$_2$(Si$_{0.8}$Ge$_{0.2}$)$_2$}
\begin{figure}[ht!]
    \centering
\includegraphics[width=0.9\linewidth]{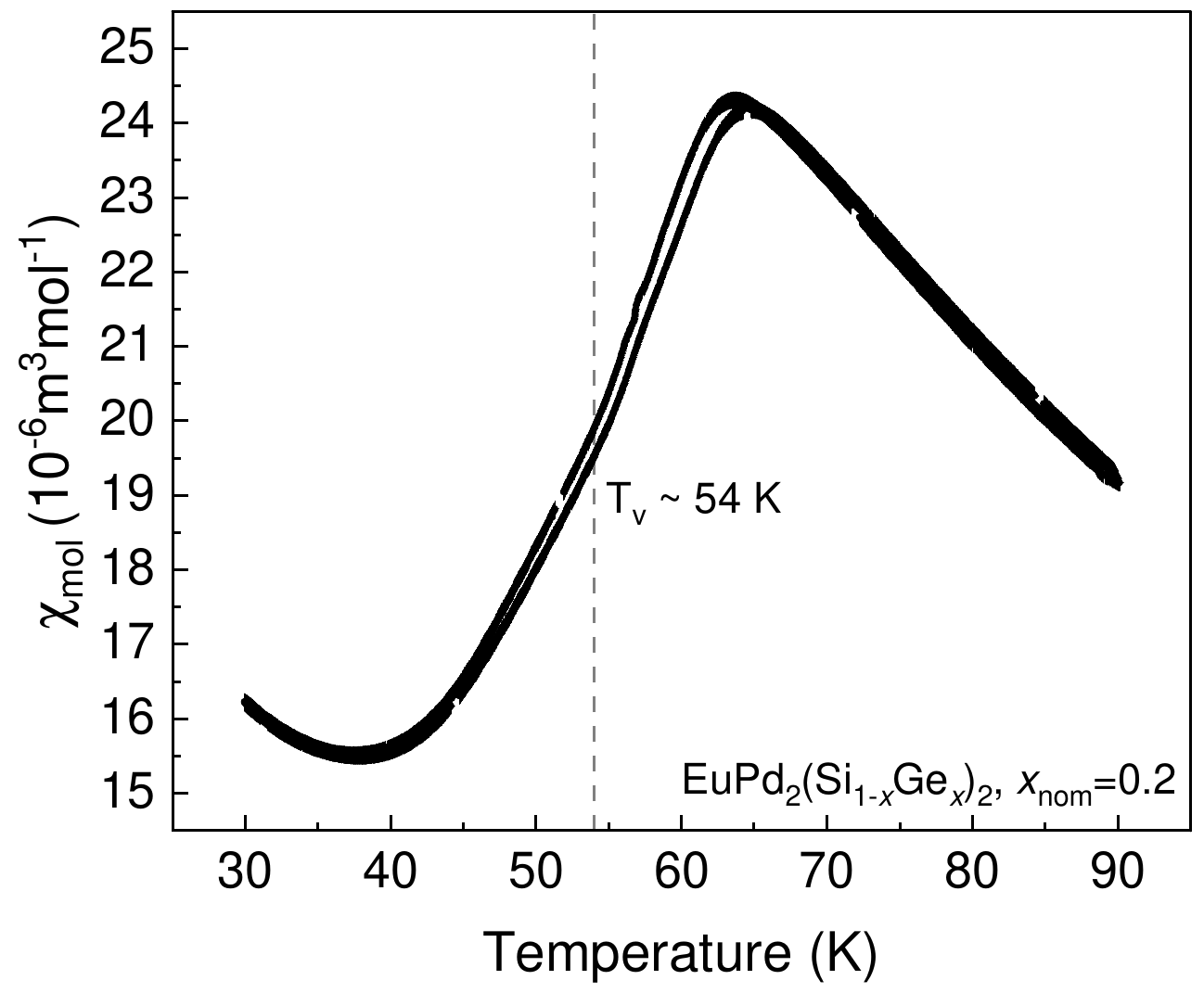}
	\caption[]{EuPd$_{2}$(Si$_{1-x}$Ge$_{x}$)$_2$: Temperature dependence of the magnetic susceptibility of a single crystalline sample with $x_{\rm nom}=0.2$. }
\label{MP809_MvT_x0p2_Ge-subst_EuPd2Si2_Tv}
\end{figure}

Measurements of magnetic susceptibility as a function of temperature are shown in Fig.~\ref{MP809_MvT_x0p2_Ge-subst_EuPd2Si2_Tv}. A crossover transition is observed at T$_V^\prime$ = 54 K. 

\end{document}